\documentclass[aps,prd,preprint,nofootinbib,groupedaddress]{revtex4}
\usepackage{amsmath,amssymb,amsbsy,color}
\usepackage{graphicx}
\usepackage{psfrag}
\usepackage{graphicx}
\newcommand{\nn}{\nonumber}

\newcommand{\pslash}{p\kern-1ex /}
\newcommand{\lslash}{l\kern-1ex /}
\newcommand{\sslash}{s\kern-1ex /}
\newcommand{\Dslash}{{\cal D}\kern-1.5ex /}

\newcommand{\tr}{{\rm tr}}
\newcommand{\vev}[1]{\langle #1 \rangle}

\newcommand{\beqa}{\begin{eqnarray}}
\newcommand{\eeqa}{\end{eqnarray}}

\newcommand{\be}{\begin{equation}}
\newcommand{\ee}{\end{equation}}
\newcommand{\bea}{\begin{eqnarray}}
\newcommand{\eea}{\end{eqnarray}}
\newcommand{\ba}{\begin{array}}
\newcommand{\ea}{\end{array}}

\newcommand{\pref}[1]{(\ref{#1})}

\newcommand{\mpis}{m_{\pi}^{2}}

\newcommand{\cowt}{\cot\omega_{\rm WT}}

\newcommand{\betamu}{\beta_{\mu}}
\newcommand{\betam}{\beta_{m}}

\newcommand{\mbare}{m_{0}}
\newcommand{\mubare}{\mu_{0}}

\begin{document}
%
\preprint{UTHEP-519}
\title{
Automatic O$(a)$ improvement for twisted-mass QCD in the presence of
spontaneous symmetry breaking
}
\author{$^{1,2}$Sinya Aoki and $^{1,3}$Oliver B\"ar}
\affiliation{
$^1$Graduate School of Pure and Applied Sciences, University of Tsukuba,
Tsukuba 305-8571, Ibaraki Japan \\
$^2$ Riken BNL Research Center, Brookhaven National Laboratory, Upton,
NY 11973, USA \\
$^3$Institute of Physics, Humboldt University Berlin, Newtonstrasse
15, 12489 Berlin, Germany
}

\date{\today}
%
\begin{abstract}
%
In this paper we
present a proof for automatic O$(a)$ improvement in
twisted mass lattice QCD at maximal twist, which uses
only  the symmetries of the leading part in the Symanzik effective  
action.
In the process of the proof we clarify
that the twist angle is dynamically determined by vacuum expectation
values in the Symanzik theory.
For maximal twist according to this definition,
we show that scaling violations of all quantities which have non-zero
values in the continuum limit are even in $a$.
In addition, using Wilson Chiral Perturbation Theory (WChPT),
we investigate this definition  for maximal twist and compare it to  
other definitions which
were already employed in actual simulations.

\end{abstract}
\maketitle
%
\section{Introduction}
%
It becomes more and more apparent that twisted mass Lattice QCD  
(tmLQCD) \cite{Frezzotti:2001ea,Frezzotti:2000nk} is a promising  
formulation to approach the chiral limit of QCD, despite the fact  
that the flavor symmetry is explicitly broken.
A twisted mass protects the Wilson-Dirac operator against small  
eigenvalues and therefore solves the problem of exceptional  
configurations, thus making numerical simulations with small quark  
masses feasible. Recent studies
\cite{Bietenholz:2004wv,Abdel-Rehim:2005gz,Jansen:2005gf,Jansen:2005kk} 
in the quenched simulation were performed with $m_{\pi}/m_ 
{\rho}$ values as small as 0.3 without running into problems due to  
exceptional configurations. Even though it will be challenging to  
reach such small pion masses in dynamical simulations, $m_{\pi}/m_ 
{\rho}< 0.5 $ seems fairly possible \cite{Urbach:2005ji}. This  
numerical advantage of tmLQCD is supplemented by the property of  
automatic O($a$) improvement at maximal twist
\cite{Frezzotti:2003ni,Frezzotti:2004wz,Frezzotti:2005zm}. For a recent review of  
these and some more results in twisted mass LQCD see Ref.\ \cite 
{Shindler:2005vj}.

Some issues, however, remain to be fully understood. The proof of  
automatic O($a$) improvement in Ref.\ \cite{Frezzotti:2003ni} makes  
use of the symmetries $m_q\equiv m_{0} - m_{\rm cr} \rightarrow -m_q
$ and $r\rightarrow -r$, where $m_{0}$ is the bare (untwisted) quark  
mass, $r$ is
the parameter in the Wilson term and $m_{\rm cr}(r)$ is the critical  
quark mass. Maximal twist is defined by setting the bare mass to a 
critical value, $m_{0}=m_{\rm cr}(r)$. A concrete definition of $m_ 
{\rm cr}(r)$ is not required in the proof
as long as the symmetry property $m_{\rm cr}(-r) = -m_{\rm cr}(r)$ is  
satisfied,
and tuning to the bare quark mass where the pion mass 
vanishes has been suggested as  one particular
choice for $m_{\rm cr}(r)$ (We call this
definition ``the pion mass definition'' in the following).
However, it has been pointed out in Ref.~\cite{Aoki:2004ta} that
the condition $m_{\rm cr}(-r) = - m_{\rm cr}(r)$ is violated for the  
pion mass
definition by non-perturbative effects.
Consequently, the O($a^{2}$) scaling violation, expected from an O($a 
$) improved theory, is lost unless the twisted quark mass satisfies  
the bound $\mu > a^{2}\Lambda_{\rm QCD}^{3}$.
Instead,  terms linear in $a$ and with fractional powers of $a$ 
are predicted by Wilson Chiral Perturbation Theory (WChPT) for very  
small twisted quark masses. On the other hand, automatic O($a$)  
improvement is expected to hold if the critical mass is defined  
through the partially conserved axial vector Ward identity quark mass  
(PCAC mass definition).

A recent paper \cite{Frezzotti:2005gi} comes to a different  
conclusion. Automatic O$(a)$ improvement has
been proven by the $m_q\rightarrow -m_q$ symmetry only, without using  
the symmetry in $r$. It
is claimed that both the pion and the PCAC mass definition guarantee  
automatic O($a$) improvement, but the remaining O($a^{2}$) effects  
differ significantly. In particular, the pion mass definition for $m_ 
{\rm cr}$ exhibits cut-off artifacts of O($a^{2}/m_{\pi}^{2})$ which  
are enhanced for small pion masses. 
These enhanced lattices artifacts are shown to be  
absent for the PCAC mass definition.

The results in Ref.\ \cite{Aoki:2004ta} and Ref.\ \cite{Frezzotti:2005gi} 
are obviously in disagreement in the small quark mass region  
where the bound $\mu > a^{2}\Lambda_{\rm QCD}^{3}$ does not hold.  
Whether this is of relevance for present day simulations with lattice  
spacings $a\approx 0.1$ fm remains to be seen. A recent scaling  
analysis \cite{Jansen:2005mf} of $f_\pi$ in quenched tmLQCD seems  
consistent with an O$(a^2)$ scaling violation, 
although its magnitude is rather large.  Given  
the fact that automatic O($a$) improvement is a highly acclaimed  
feature of twisted mass LQCD, it is certainly desirable to study this  
issue further and find an explanation for these contradicting results.

Closely related is the so-called 'bending phenomenon', observed in  
quenched simulations
\cite{Bietenholz:2004wv,Abdel-Rehim:2005gz,Jansen:2005gf,Jansen:2005kk}. 
The pion mass, the pion decay constant and the vector meson  
mass show an unexpected strong  non-linear quark mass dependence for  
small quark masses if the pion mass definition for the critical quark  
mass is used.
This curvature is significantly reduced when the untwisted quark mass  
is tuned according to the optimal choice proposed in \cite{Frezzotti:2005gi}.
That this bending is indeed a lattice artifact of the twisted mass  
formulation is also supported by calculations using the overlap  
operator on the same gauge field configurations with similarly small  
pion masses. Here the bending is absent \cite{Bietenholz:2004wv}.

In this paper we revisit the property of automatic O($a$) improvement  
in twisted mass QCD.
We first give an alternative proof for automatic O$(a)$ improvement
at maximal twist without using the symmetries of the parameters $m_q$  
and $r$.
Although our proof is just an improved version of previous
ones \cite{Frezzotti:2003ni,Frezzotti:2005gi}, we can clarify the
meaning of  ``maximal twist'' in the process of our proof. We will  
argue that, in presence of spontaneous symmetry breaking, the twist  
angle $\theta$ is determined dynamically by the ratio of two vacuum  
expectation values in the Symanzik theory, namely
\beqa
\cot \theta &=&\frac{\langle \bar\psi\psi\rangle}{\langle \bar\psi i
\gamma_5
\tau^3 \psi\rangle}.
\eeqa
Provided that the mass parameters of the theory are tuned such that $ 
\theta = \pm\pi/2$, we can show that the scaling violations of  
observables start with $a^{2}$, i.e. the theory is O$(a)$ improved.

We also investigate
this new definition for maximal twist using Wilson Chiral  
Perturbation Theory (WChPT) \cite{Sharpe:1998xm,Rupak:2002sm} (for a  
review see Ref.\ \cite{Bar:2004xp}).
We explicitly show the absence of O($a,a\mu)$ contributions in the  
expressions for the pion mass and decay constant.

We finally compare   
our new criterion with other definitions of maximal twist, the pion mass and the PCAC mass definition,
which were previously employed in numerical simulations.  We find  
that, although these two definitions show
asymptotic $a^2$ scaling violations, they do not exhibit the expected $a^2$
scaling until $a$ becomes small such that the bound $\mu > a^{2} 
\Lambda_{\rm QCD}^{3}$ is satisfied. 

%
\section{Alternative Proof of O$(a)$ improvement}
\label{sec:proof}
%
\subsection{Main idea for O$(a)$ improvement}
\label{subsec:idea}
%
The twisted mass lattice QCD action for the 2-flavor theory is given by
\beqa
S_{\rm tmQCD} &=& S_G + S_{\rm tm}.
\label{eq:action}
\eeqa
The details of the gauge action $S_G$ are irrelevant in the
following, so we leave it unspecified.
$S_{\rm tm}$ denotes the 2-flavor Wilson fermion action with a
twisted  mass term,
which is defined as
\beqa
S_{\rm tm} &=& \sum_{x,\mu} \bar \psi_L (x) \frac{1}{2}\left[ \gamma_\mu
(\nabla_\mu^+ + \nabla_\mu^-)\psi_L
-ar \nabla_\mu^+ \nabla_\mu^-\psi_L\right](x) +
\sum_x\bar\psi_L (x) M_0 e^{i\theta_{0} \gamma_5\tau^3} \psi_L (x)
\label{eq:lattice_action}
\eeqa
with
\beqa
\left[\nabla_\mu^+ \psi_L\right](x) &=& \frac{1}{a}
\left(U_\mu (x)\psi_L(x+\mu)-\psi_L (x)\right), \\
\left[\nabla_\mu^- \psi_L\right](x) &=& \frac{1}{a}\left(
\psi_L (x)-U_\mu^\dagger (x-\mu)\psi_L (x-\mu)\right),
\eeqa
being the standard forward and backward difference operators. We
supplemented the fields with the subscript ''{\em L}'' in order to
highlight the fact that these fields are lattice fields. The
parameters $M_{0}$ and $\theta_{0}$ denote the bare mass and bare
twist angle. Instead of using this exponential notation it is also
customary to write
\bea\label{tmstandardRep}
M_0 e^{i\theta_{0} \gamma_5\tau^3} &= & m_0 + i \mu_{0}\gamma_5\tau^3,
\eea
where the bare untwisted mass $m_{0}$ and the bare twisted mass $\mu_
{0}$ are given by
\bea
m_{0} & = & M_0 \cos\theta_{0},\qquad\quad
\mu_{0} \, = \, M_0 \sin\theta_{0}.
\eea

The  lattice action (\ref{eq:action}) is invariant under the
following global symmetry transformations
\cite{Frezzotti:2000nk,Sharpe:2004ps}:
\begin{enumerate}
\item U(1)$\otimes$ U(1) vector symmetry
\[
\psi_L \rightarrow e^{i (\alpha_0 + \alpha_3\tau^3)}\psi_L ,\quad
\bar\psi_L \rightarrow \bar\psi_L e^{-i (\alpha_0 + \alpha_3\tau^3)} .
\]
This transformation is part of the U(2) flavor symmetry of the
untwisted theory.
\item Extended parity symmetry
\beqa
P_F^{1,2}:\  \psi_L &\rightarrow&  \tau^{1,2} P\psi_L, \quad
\bar\psi_L \rightarrow P \bar\psi_L \tau^{1,2},
\label{eq:exParity}
\eeqa
where $P$ is the parity transformation, given by
\[
P\psi_L(\vec{x},t) = i \gamma_0 \psi_L(-\vec{x},t), \quad
P\bar\psi_L(\vec{x},t) =-i\bar\psi_L(-\vec{x},t) \gamma_0 .
\]
For the gauge fields $P_F$ is equal to the standard parity
transformation. Note that ordinary parity $P$ is not a symmetry
unless it is combined with a flavor rotation in the 1 or 2 direction.
Alternatively, one can also augment $P$ with a sign change of the
twisted mass term $\mu_{0}$,
\bea\label{paritytrafo2}
\tilde{P} & =& P \times [\mu_{0} \rightarrow -\mu_{0}],
\eea
which is also a symmetry of the action.

\item Charge conjugation symmetry
\[
C:\ \psi_L(x) \rightarrow i\gamma_0\gamma_2 \bar\psi_L(x)^T,\quad
\bar\psi_L(x) \rightarrow -\psi_L(x)^T i\gamma_0\gamma_2 ,
\]
together with the charge conjugation transformation for the gauge
fields, $U(x,\mu) \rightarrow U(x,\mu)^{*}$ .
\end{enumerate}
Besides these symmetries the lattice action is also invariant under
hypercubic lattice rotations and local gauge transformations.

According to Symanzik \cite{Symanzik:1983dc,Symanzik:1983gh}, the
lattice theory can be described by an effective continuum theory.
The form of the effective action of this theory is restricted by
locality and  the symmetries of the underlying lattice theory. Taking
into account the symmetries listed above one finds \cite{Sharpe:2004ps}
\bea
S_{\rm eff} &=& S_0 + a S_1 + a^{2} S_{2} + \cdots,
\eea
where the first two terms are given as
\bea
S_{0} &=& S_{0,\rm gauge} + \int d^4x\, \bar\psi(x) \left[ \gamma_
{\mu}D_{\mu} +
M_R e^{i\theta\gamma_5\tau^3}\right]\psi(x),\label{eq:symanzik_S0}\\
S_{1} & = & C_1
\int d^4x\, \bar\psi(x) \sigma_{\mu\nu} F_{\mu\nu}(x) \psi (x).
\label{eq:symanzik_S1}
\eeqa
$S_{0,\rm gauge}$ denotes the standard continuum gauge field action
in terms of the gauge field tensor $F_{\mu\nu}$. The second term in
$S_{0}$ is the continuum twisted mass fermion action. The mass
parameters are renormalized masses, and we assume the renormalization
scheme in \cite{Frezzotti:2001ea}.\footnote{Other choices for the
renormalized parameters are of course possible, but at the expense of
additional terms in $S_{1}$ of the effective action 
\cite{Sharpe:2004ps}.
We also assume that use of the leading order equations of
motion has been made in order to drop some O($a$) terms in $S_{1}$.
Without using the renormalization scheme in \cite{Frezzotti:2001ea}
and equations of motion there would be seven terms present in $S_{1}$
instead of only one \cite{Sharpe:2004ps}.
However, this larger number of terms would not alter the conclusion of this section.
} It is worth mentioning that there
is no ``twisted'' Pauli term $\bar\psi \gamma_5\tau^3 \sigma_{\mu\nu}  
F_{\mu\nu}
\psi$ present in $S_{1}$, since such a term violates the symmetry in
eq.\ \pref{paritytrafo2}.\footnote{This property can also be derived  
from a different point of view.
Since parity is conserved at $\mu_0=0$ in the lattice theory,
$\bar\psi \gamma_5\tau^3 \sigma\cdot F\psi$
does not appear without $\mu_0$. This argument can be extended to the
case where the parity-flavor symmetry is spontaneously broken  for a
certain range of the untwisted mass $M_0\cos\theta_0$
\cite{Aoki:1983qi,Aoki:1985mk,Aoki:1986kt,Aoki:1986xr,Aoki:1987us}
in the lattice theory.  In this case, the charged pions
become massless Nambu-Goldstone bosons in the lattice
theory, associated with this spontaneous symmetry breaking in the
zero twisted mass limit.
Therefore, it must also become massless in the Symanzik theory in the
same limit.
This fact also tells us that explicit parity-flavor breaking terms
such as
$\bar\psi i\gamma_5\tau^3 \sigma_{\mu\nu} F_{\mu\nu}\psi$ must be  
absent in
the Symanzik theory without $\mu_0$.
}

In addition to the effective action we have to specify the direction
of the chiral condensate, since chiral symmetry is spontaneously
broken. From the fact that the direction of the chiral condensate is
completely controlled by the direction of the symmetry breaking
external field
(i.e. the quark mass) in the continuum theory, we can take
\beqa
\langle \bar\psi_\alpha^i\psi_\beta^j\rangle_{S_0}
&=& \frac{v(M_R)}{8}\left[ e^{-i\theta\gamma_5\tau^3}\right
]_{\beta\alpha}^{ji},
\label{eq:vev0}
\eeqa
where $\lim_{M_R\rightarrow 0} \lim_{V \rightarrow\infty} v(M_R)\not=
0$.\footnote{The computation of this condensate follows standard
arguments where one first considers the theory in a finite box with 4-
volume $V$. See, for example, the appendix of Ref.\ \cite{Sharpe:1998xm}.}
Here the vacuum expectation value (VEV) is defined with respect to 
the continuum action $S_0$.
To say it differently, the VEV (\ref{eq:vev0}) defines the twist
angle $\theta$ in the Symanzik theory.
The above condensate is equivalent to
\beqa
\langle \bar\psi \psi\rangle_{S_0} &=& v(M_R)\cos \theta,\\
\langle \bar\psi i\gamma_5\tau^3 \psi\rangle_{S_0} &=& v(M_R)\,
\sin \theta .
\eeqa

We now want to argue
that the choice $\theta = \pi/2$ (or $-\pi/2$) corresponds to
maximal twist. In terms of the mass parameters this is equivalent
to $M_{R} = \mu_{R}$ and $m_{R}=0$. In this case the action and the
VEVs become
\bea
S_0 &=&S_{0, \rm gauge}+ \int d^4x\ \bar\psi \left[
\gamma_{\mu} D_{\mu}+ i M_R  \gamma_5\tau^3 \right] \psi (x), \label{eq:symanzik2}\\
S_1 & = &  C_1 \int d^4x\ \bar\psi(x) \sigma_{\mu\nu} F_{\mu\nu}(x)
\psi(x) ,
\eea
\bea
\langle \bar\psi \psi\rangle_{S_0} &=& 0,
\label{eq:p0}\\
\langle \bar\psi i\gamma_5\tau^3 \psi\rangle_{S_0} &=&  v(M_R) ,
\label{eq:p5}
\eea
It is easy to check that $S_0$,
the continuum part of the effective action
is invariant under
\beqa
\psi \rightarrow e^{i w\gamma_5\tau^{1,2}}\psi, \qquad
\bar\psi \rightarrow \bar\psi e^{i w\gamma_5\tau^{1,2}},
\label{eq:exsym}
\eeqa
and therefore also under the $Z_2$ subgroup $T_1$ of this continuous
transformation,
defined by\footnote{A similar argument using this symmetry has been
given independently by S. Sint \cite{SintNara}.}
\beqa
T_1 \psi &=& i\gamma_5 \tau^1 \psi, \qquad
T_1\bar\psi =  \bar\psi i\gamma_5 \tau^1.
\label{eq:T1}
\eeqa
Since $T_1^2 =1$ in the space of fermion number conserving operators,
which contain equal numbers of $\psi$ and $\bar\psi$,
the eigenvalues of $T_1$ are $1$ ($T_{1}$-even) or $-1$ ($T_{1}$-odd).
The crucial observation is that the VEVs (\ref{eq:p0}) and (\ref
{eq:p5}) are also invariant
under this transformation. The symmetry (\ref{eq:exsym}) (and its
discrete subgroup $T_1$) is {\em not}  spontaneously broken, hence it
is an exact symmetry of the continuum theory.
The O$(a)$ term
\beqa
a S_1 &=& a C_1\int d^4 x\
\bar\psi(x) \sigma_{\mu\nu} F_{\mu\nu}(x) \psi (x) ,
\eeqa
on the other hand, is odd under $T_1$.
Therefore non-vanishing physical observables, which
must be even under $T_1$, can not have an O$(a)$ contribution, since
the O$(a)$ term is odd under $T_1$ and therefore must vanish
identically.
This is automatic O$(a)$ improvement at maximal twist.\footnote{This
argument does not rely on our particular renormalization scheme and
the use of the equations of motion. All possible terms in $S_{1}$ are
$T_{1}$-odd once the continuum part is invariant under the
transformation \pref{eq:T1}.}
Note that non-invariant, i.e.\ $T_1$-odd quantities, which vanish in
the continuum limit, can have O$(a)$ contributions.

The above argument gives just the main idea of our proof for
automatic O$(a)$ improvement, and we will give a detailed proof in
the next
subsection. However,
one of the most important points of our analysis is that the
condition for maximal twist and for automatic O$(a)$
improvement is determined dynamically by the VEV $ \vev{\bar\psi_
\alpha^i \psi_
\beta^j}$ in the Symanzik theory.
More explicitly, the symmetry (\ref{eq:exsym})
or (\ref{eq:T1}) of the continuum theory must keep the VEV intact,
so that the symmetry is not spontaneously broken.
This condition seems
natural, since the symmetry (\ref{eq:exsym})
corresponds to a part of the exact vector symmetry in the
continuum QCD at maximal twist. 
Note that this symmetry refers to the vector symmetry 
in the so-called 
twisted basis \cite{Frezzotti:2000nk,Frezzotti:2003ni}. After
rotating to the physical basis the theory is invariant under ordinary
vector rotations in the 1 or 2 direction. However, for the proof of O($a$)
improvement in the next subsection we prefer to stay in the
twisted basis.

%
\subsection{Proof of O$(a)$ improvement}
\label{subsec:proof}
%
Let us consider an arbitrary  multi-local lattice operator ${\cal O}_
{\rm lat}^{tp,d}( \{x\} )$,
where $\{x\} $ represents $x_1,x_2, \cdots, x_n$,  $d$ is the canonical
dimension
of the operator,  $t=0,1$ and $p=0,1$ denote transformation
properties under $T_1$ and parity $P$:
\beqa
T_1 :\  {\cal O}_{\rm lat}^{tp,d} (\{x\}) &\rightarrow& (-1)^t{\cal O}
_{\rm lat}^{tp,d} (\{x\}) , \\
P : \ {\cal O}_{\rm lat}^{tp,d} (\{\vec{x}, t\}) &\rightarrow& (-1)^p
{\cal O}_{\rm lat}^{tp,d} (\{ -\vec{x},t \}) .
\eeqa
Here we do not include the dimension coming from powers of the quark  
mass  in the canonical
dimension $d$ of operators. For example,
\beqa
{\cal O}_{\rm lat}^{01,3}(x) &=& \bar\psi_L i \gamma_5\tau^3\psi_L
(x), \quad
{\cal O}_{\rm lat}^{10,3}(x) = \bar\psi_L \psi_L (x)
\label{eq:op013}, \nn \\
{\cal O}_{\rm lat}^{00,4}(x) &=& \sum_\mu \bar\psi_L\frac{1}{2}\gamma_
\mu
\left(\nabla_\mu^++\nabla_\mu^-\right) \psi_L (x),  \nn \\
{\cal O}_{\rm lat}^{10,5}(x) &=& \sum_\mu \bar\psi_L\frac{1}{2}\nabla_
\mu^+ \nabla_\mu^- \psi_L (x),
\label{eq:op105}
\eeqa
and in terms of these operators the lattice action is given by
\beqa
S_{\rm tm} &=& \sum_x \left[ {\cal O}_{\rm lat}^{00,4}(x) - ar\ {\cal
O}_{\rm lat}^{10,5}(x)
+m_0\  {\cal O}_{\rm lat}^{10,3}(x)+ \mu_0\ {\cal O}_{\rm lat}^{01,3}(x)
\right],
\label{eq:OP_action}
\eeqa
with untwisted quark mass $m_0=M_0\cos\theta_0$ and twisted quark mass $
\mu_0 = M_0\sin\theta_0$.

The lattice operator ${\cal O}_{\rm lat}^{tp,d}$ corresponds to a sum of
continuum operators ${\cal O}^{t_n p_n,n}_{i_n}$  ($n$: non-negative
integer) in the Symanzik theory as
\beqa
{\cal O}_{\rm lat}^{tp,d} &\Leftrightarrow & {\cal O}_{\rm eff}^{tp,d}
=\sum_{n=0}^\infty a^{n-d} \sum_{t_n,p_n}\sum_{i_n} c_{t_n p_n,n,i_n}^
{tp,d} {\cal O}
^{t_np_n,n}_{i_n}  ,
\label{eq:OP_expand}
\eeqa
where $n$ is the canonical dimension of the continuum operator ${\cal O}
^{t_np_n,n}_{i_n}$
which consists of  $\bar\psi$, $\psi$, $A_\mu$ and $D_\mu$ only
without any mass parameters, and
\beqa
T_1 : \  {\cal O}^{t_np_n,n}_{i_n} (\{x\})&\rightarrow & (-1)^{t_n}
{\cal O}
^{t_np_n,n}_{i_n}(\{x\}),\\
P : \  {\cal O}^{t_np_n,n}_{i_n}(\{ \vec{x},t \}) &\rightarrow & (-1)^
{p_n}
{\cal O}^{t_np_n,n}_{i_n} (\{ -\vec{x},t \}),
\eeqa
with $t_n, p_n =0,1$. Here we distinguish different operators with
the same $(t_n p_n, n)$ by an index $i_n$.
To have a total dimension $d$ in the expansion in Eq.\ (\ref 
{eq:OP_expand}),
the coefficients $c_{t_np_n,n,i_n}^{tp,d}$ must be dimensionless:
$c_{t_np_n,n,i_n}^{tp,d} = c_{t_np_n,n,i_n}^{tp,d}(g^2, \log(\Lambda a),
m_qa, \mu_0 a)$,
where $g^2$ is the bare gauge coupling constant, $\log (\Lambda a)$
represents log-divergences of
the lattice theory with some scale parameter $\Lambda$ introduced in
the Symanzik theory,
and $m_q = m_0 - m_{\rm cr}$ is a subtracted quark mass with an  
additive mass
counter term $m_{\rm cr}$,
which will be specified later.
Note that we consider possible power divergences of lattice operators by
including operators with $n=0 ,1,\cdots, d-1$ in the expansion.

The following selection rules among these operators
are crucial for our proof of automatic O$(a)$ improvement:
\beqa
t  + p+d &=& t_n  + p_n +n\quad {\rm mod} (2) ,
\label{eq:rule1} \\
p+ \#\mu_0 &=& p_n + (\#\mu_0)_n \quad {\rm mod} (2),
\label{eq:rule2}
\eeqa
where  $\#\mu_0$ and $(\#\mu_0)_n$ denote the numbers of $\mu_0$'s in $
{\cal O}_
{\rm lat}^{tp,d}$ and
$c_{t_np_n,n,i_n}^{tp,d}$, respectively.
The second equality can be easily proven by the invariance of the
lattice action
(\ref{eq:lattice_action}) under the $\tilde P =  P \times [\mu_0
\rightarrow -\mu_0 ]$
transformation, Eq. (\ref{paritytrafo2}).
To prove the first equality (\ref{eq:rule1}), we introduce
the following transformation:
\beqa
{\cal D}_d^1 : \left\{
\begin{array}{ccc}
U_\mu(x) & \rightarrow & U^\dagger_\mu(-x-a\mu) \\
\Bigl(A_\mu (x) &\rightarrow & - A_\mu(-x) \Bigr)\\
\psi(x) &\rightarrow &\left( e^{i\pi\tau_1}\right)^{3/2}\psi(-x) \\
\bar\psi(x) &\rightarrow & \bar\psi(-x) \left(e^{i\pi\tau_1}\right)^
{3/2} \\
\end{array}
\right. ,
\label{eq:Dd1}
\eeqa
which is a modified version of the transformation ${\cal D}_d$
introduced in Ref.~\cite{Frezzotti:2003ni}.
Since
it is easy to show that
the lattice action (\ref{eq:lattice_action}) is invariant under $T_1
\times {\cal D}_d^1$, in addition to the invariance under $P_F^1$,
the lattice action is invariant under $T_1 \times {\cal D}_d^1\times
P_F^1$.
On the other hand,
combining the transformation  property
\beqa
{\cal D}_d^1 : \left\{
\begin{array}{ccc}
\left(\nabla_\mu^+ + \nabla_\mu^-\right)\cdot f(x) & \rightarrow &
- \left( \nabla_\mu^+ + \nabla_\mu^- \right)\cdot f(-x) \\
\Bigl(D_\mu \cdot f(x) &\rightarrow & - D_\mu\cdot f(-x) \Bigr)\\
\end{array}
\right.
\eeqa
for an arbitrary function $f(x)$ with eqs.(\ref{eq:Dd1}) and (\ref
{eq:exParity}),
we can easily see that
${\cal D}_d^1\times P_F^1$ counts the canonical dimension times the parity
of the operator as
\beqa
{\cal D}_d^1 \times P_F^1 : \ {\cal O}_{\rm lat}^{tp,d}( \{\vec{x},t\})
&\rightarrow & (-1)^{d+p}
    {\cal O}_{\rm lat}^{tp,d}( \{\vec{x},-t\}) \\
{\cal D}_d^1 \times P_F^1 : \ {\cal O}^{t_np_n,n}_{i_n}( \{\vec{x},t\}) &
\rightarrow & (-1)^{n+ p_n}
{\cal O}^{t_np_n,n}_{i_n}( \{\vec{x},-t\}) .
\eeqa
Therefore, the invariance of the action under $T_1\times {\cal D}_d^1
\times P_F^1$ implies the first equality (\ref{eq:rule1}).

Let us show how these selection rules are used to determine the
structure of operators in the Symanzik theory.
As an example we consider the operator ${\cal O}^{01,3}_{\rm lat}(x)$
in eq.(\ref{eq:op013}). Since $t+p+d=4$, the first selection rule gives
$t_n+d_n+n = 0$ ${\rm mod}(2)$,
which leads to
\beqa
{\cal O}^{01,3}_{\rm eff} &=& \frac{c^{01,3}_{00,0}}{a^3}{\cal O}^
{00,0}+
c^{01,3}_{10,3}{\cal O}^{10,3}+c^{01,3}_{01,3}{\cal O}^{01,3}+
a\, c^{01,3}_{00,4,A}{\cal O}^{00,4}_A+ a\, c^{01,3}_{00,4,B}{\cal O}^
{00,4}_B+
a\, c^{01,3}_{11,4}{\cal O}^{11,4} \nn \\
&+& a^2\, c^{01,3}_{01,5,A}{\cal O}^{01,5}_A+a^2\, c^{01,3}_{01,5,B}
{\cal O}^{01,5}_B+
a^2\, c^{01,3}_{10,5,A}{\cal O}^{10,5}_A+a^2\, c^{01,3}_{10,5,B}{\cal
O}^{10,5}_B+
\cdots
\eeqa
where
\beqa
{\cal O}^{00,0} &=& {\bf 1}, \quad
{\cal O}^{01,3}= \bar\psi i \gamma_5\tau^3\psi , \quad
{\cal O}^{10,3} = \bar\psi \psi, \nn \\
{\cal O}^{00,4}_A &=& \bar\psi\gamma_\mu D_{\mu}  \psi, \quad
{\cal O}^{00,4}_B = -\frac{1}{4}\tr F_{\mu\nu}F_{\mu\nu}, \quad
{\cal O}^{11,4} = \bar\psi\gamma_\mu D_{\mu} i\gamma_5\tau^3  \psi,
\nn \\
{\cal O}^{01,5}_A &=& \bar\psi i\gamma_5\tau^3 D^2  \psi, \
{\cal O}^{01,5}_B = \bar\psi i\gamma_5\tau^3 \sigma_{\mu\nu}F_{\mu
\nu} \psi, \
{\cal O}^{10,5}_A = \bar\psi D^2  \psi, \
{\cal O}^{10,5}_B = \bar\psi \sigma_{\mu\nu}F_{\mu\nu} \psi .
\eeqa
Applying the second selection rule that $p+\#\mu_0 = 1 = p_n + (\#\mu_0)
_n$ ${\rm mod}(2)$,
we obtain
\beqa
c^{01,3}_{00,0}(\mu_0 a)&=& \mu_0 a\, \tilde c^{01,3}_{00,0}(\mu_0^2
a^2), \quad
c^{01,3}_{10,3} (\mu_0 a)= \mu_0 a\, \tilde c^{01,3}_{10,3}(\mu_0^2
a^2), \nn \\
c^{01,3}_{00,4,A(B)}(\mu_0 a) &=& \mu_0 a\, \tilde c^{01,3}_{00,4,A
(B)}(\mu_0^2 a^2), \quad
c^{01,3}_{10,5,A(B)}(\mu_0 a) = \mu_0 a\, \tilde c^{01,3}_{10,5,A(B)}
(\mu_0^2 a^2),
\eeqa
where only the $\mu_0 a$ dependence is explicitly written,
and the other $c^{01,3}_{t_np_n,n,i_n}$'s are even functions of $ 
\mu_0 a$.
We then finally have
\beqa
{\cal O}^{01,3}_{\rm eff} &=& a^{-2}\mu_0\, \tilde c^{01,3}_{00,0}
{\cal O}^{00,0}+
c^{01,3}_{01,3}{\cal O}^{01,3}+
a^2\, \mu_0\, \tilde c^{01,3}_{00,4,A}{\cal O}^{00,4}_A+ a^2\,\mu_0\,
\tilde c^{01,3}_{00,4,B}{\cal O}^{00,4}_B\nn \\
&+&
a^2\, c^{01,3}_{01,5,A}{\cal O}^{01,5}_A+a^2\, c^{01,3}_{01,5,B}{\cal
O}^{01,5}_B\nn \\
&+& a\, \mu_0\, \tilde c^{01,3}_{10,3}{\cal O}^{10,3}+
a\, c^{01,3}_{11,4}{\cal O}^{11,4} +
a^3\,\mu_0\, \tilde c^{01,3}_{10,5,A}{\cal O}^{10,5}_A+a^3\mu_0\,
\tilde c^{01,3}_{10,5,B}{\cal O}^{10,5}_B+
\cdots,
\eeqa
where all dimensionless functions are even in $\mu_0 a$.
It is important to observe that all operators with $t=0$
contain only  even powers of $a$, while those with $t=1$ have only  
odd powers
of $a$.

Repeating the analysis given above for all operators which appear in the
lattice action (\ref{eq:OP_action})
and introducing renormalized quantities (see appendix \ref{appendixB}
for more details),
we obtain
\beqa
S_{\rm tmQCD} \Leftrightarrow S_{\rm eff} &=& S_0  + m_q S_m +\sum_
{n=1}^\infty\left[
a^{2n}S_{2n}^0 + a^{2n-1} S_{2n-1}^1\right],
\label{eq:SymAction}
\eeqa
where
\beqa
S_0 &=&\int d^4 x \left\{ \bar\psi_R \left( \gamma_{\mu} D_{\mu} + i  
\mu_R
\gamma_5\tau^3 \right)\psi_R (x)
-\frac{1}{4}F_{\mu\nu,R}^2(x)\right\},\label{eq:SymAction0}
\\
m_q S_m &=& m_R S_{m_R} \equiv m_R \int d^4 x  \bar\psi_R\psi_R (x),
\label{eq:SymActionMass}\\
S_{2n}^0 &=& \int d^4x\ \left[\sum_{i} Z_{00,2n+4}^{i}\cdot {\cal O}_
{R.i}^{00,2n+4}
(x)+\sum_i Z_{01,2n+3}^i\cdot \mu_R
\cdot {\cal O}_{R,i}^{01,2n+3}(x)\right],
\label{eq:SymActionEven} \\
S_{2n-1}^1 &=& \int d^4x\ \left[\sum_i  Z_{10,2n+3}^i\cdot {\cal O}_
{R,i}^{10,2n+3}
(x)+\sum_i Z_{11,2n+2}^i\cdot
\mu_R \cdot {\cal O}_{R,i}^{11,2n+2}(x)\right].
\label{eq:SymActionOdd}
\eeqa
Renormalized parameters are introduced as\footnote{Note that the  
renormalization differs from the one usually employed in the Symanzik  
improvement program.}
\beqa
\mu_0 &=& Z_\mu^{-1} (g_R,\log(\Lambda a), m_R a, \mu_R^2 a^2)
\mu_R,\quad
m_q = Z_m^{-1} (g_R,\log(\Lambda a), m_R a, \mu_R^2 a^2) m_R,
\nonumber \\
g &=& Z_G^{1/2} (g_R,\log(\Lambda a), m_R a, \mu_R^2 a^2) g_R,
\label{eq:redefine_para}
\eeqa
where $g_R, m_R, \mu_R$ are kept constant and finite
as $a\rightarrow 0$.
We also define renormalized fields as
\beqa
\psi_R &=& Z_F^{1/2}(g_R^2, \log(\Lambda a), m_Ra, \mu_R^2 a^2)
\psi, \quad
A_{\mu,R} = Z_G^{1/2} (g_R^2, \log(\Lambda a), m_Ra, \mu_R^2 a^2) A_
\mu .
\label{eq:redefine_field}
\eeqa
A subscript $R$ in ${\cal O}_{R,i}^{t_n p_n,n}$  means that the  
operators
are expressed in terms of renormalized fields,
and therefore $Z^i_{t_np_n,n}= Z^i_{t_np_n,n}(g_R^2,\log(\Lambda a),  
m_R a,
\mu_R^2 a^2)$.  

Similarly, applying the selection rules to an arbitrary operator
${\cal O}_{\rm lat}^{tp,d}$ (again we give more details in appendix  
\ref{appendixB}), we obtain
\beqa
{\cal O}_{\rm lat}^{tp,d} &\Leftrightarrow & {\cal O}^{tp,d}_{\rm eff}=
\sum_{l=-d}^\infty a^{l} \left[ \sum_i c^{tp,d}_{[t+l]p,d+l,i}\,
{\cal O}_{R,i}^{[t+l]p,d+l}
+ \mu_R\,\sum_i
\tilde c^{tp,d}_{[t+l]\bar p,d+l-1,i}\, {\cal O}_{R,i}^{[t+l]\bar p. d
+l-1}\right],
\label{eq:OP_expand2}
   \eeqa
where $[t+l]=t+l$ mod(2), $\bar p = 1- p$, and
coefficients $c_{t_n p,n,i}^{tp,d}$ and
$\tilde c_{t_n p,n.i}^{tp, d}$ are even functions of $\mu_R a$.
Note here that, even though we use the same notations as in eq. (\ref
{eq:OP_expand}),
these coefficients are functions of $g_R^2, \log(\Lambda a), m_Ra$
and $\mu_R^2 a^2$,
therefore the functional forms are different from the original ones.
Formula \pref{eq:OP_expand2} tells us that, if the lattice operator  
has $t=0$,
operators with $t_n=0$ in the Symanzik
expansion appear  with even powers of $a$ while those with $t_n=1$
are associated with odd
powers of $a$. This is reversed in the case that  the lattice
operator has $t=1$:
operators with $t_n=0$ are multiplied by odd powers of $a$ in the
Symanzik expansion
and those with $t_n=1$ by even powers of $a$.

In order to obtain a finite result in the continuum limit,
we have to remove possible power divergences in the expansion (\ref
{eq:OP_expand2})
by subtracting lower dimensional lattice operators from the original
operator ${\cal O}_{\rm lat}$,
in addition to subtractions of $\log (\Lambda a)$ divergences
including a possible mixing among
operators whose canonical dimension is same as the original operator.
We denote such a renormalized and subtracted operator as ${\cal O}_
{\rm lat, sub}$,
which corresponds to
\beqa
{\cal O}_{\rm lat,R,sub}^{tp,d} \Leftrightarrow  {\cal O}^{tp,d}_{\rm
eff,R,sub}&=&{\cal O}_{R}^{tp,d} +\sum_{l=1}^\infty a^{l}
{\cal O}^{[t+l], d+l}_{R;tpd}, \label{eq:OP_expand3}\\
{\cal O}^{[t+l],d+l}_{R; tpd}
&=&
\sum_i c^{tp,d}_{[t+l]p,d+l,i}\, {\cal O}_{R,i}^{[t+l]p,d+l}
+ \mu_R\, \sum_i
\tilde c^{tp,d}_{[t+l]\bar p,d+l-1,i}\, {\cal O}_{R,i}^{[t+l]\bar p. d
+l-1}\  ,
\eeqa
where $d+l$ in the
short-hand notation ${\cal O}^{[t+l],d+l}_{R;tpd}$
represents a
canonical dimension of the operator and $[t+l]$ labels the
transformation property under
$T_1$:
\beqa
T_1:\ {\cal O}^{[t+l],d+l}_{R;tpd} &\rightarrow& (-1)^{t+l}  {\cal O}^
{[t+l],d+l}_{R;tpd}.
\eeqa

We conclude that, in terms of this general description in the
Symanzik theory, the maximal twist condition corresponds to the  
property that the
continuum theory is invariant under the $T_1$ transformation.
This condition then entails $m_R = 0$, which we call exact invariance  
condition. However, it can be relaxed to $m_R={\rm O}(a)$, which we  
call weak invariance condition. Imposing either of these we find
\beqa
\langle {\cal O}_R^{1p,d} \rangle_{S_0+ m_R S_{m_R}} &=& \frac{1}{Z}
\int {\cal D}\psi_R {\cal D}\bar\psi_R {\cal D}A_{\mu, R}e^{S_0 +m_R
S_{m_R}}\
{\cal O}_R^{1p,d} =\left\{ \begin{array}{cc}
0 &\mbox{  exact  } \\
   {\rm O}(a) & \mbox{  weak  }\end{array}\right.
\label{eq:maximal}
\eeqa
for an arbitrary continuum operator ${\cal O}_R^{1p,d}$ which is odd  
under $T_1$.
(In the operator formalism, this condition expresses the fact that  
the vacuum $
\vert 0\rangle$ of $S_0 + m_R S_{m_R}$ is
invariant under $T_1$:  $ {\cal T}_1 \vert 0 \rangle = 0$ or O$(a)$.)

Assuming the maximal twist condition is satisfied, i.e.\ $m_R = O(a)$ at
least, we now consider the following correlation function:
\beqa
\langle {\cal O}_{\rm lat,R,sub}^{tp,d}( \{ x \} ) \rangle &\equiv &
\frac{1}{Z_{\rm lat} }
\int {\cal D}\psi_L {\cal D}\bar\psi_L {\cal D}U \
e^{S_{\rm tmQCD}}\  {\cal O}_{\rm lat,R,sub}^{tp,d}( \{ x \} )
\eeqa
where
$Z_{\rm lat}$ is the partition function defined by $\langle 1 \rangle  
=1$.
In terms of the Symanzik effective theory, this correlation function
corresponds to
\beqa
\langle {\cal O}_{\rm lat, R,sub}^{tp,d}( \{ x\} )  \rangle &=&
\langle {\cal O}_{\rm eff, R,sub}^{tp,d}( \{ x\} )  \rangle_{S_{\rm
eff}}
\eeqa
where we define
\beqa
\langle {\cal O} \rangle_S &=& \frac{1}{Z}
\int {\cal D}\psi_R {\cal D}\bar\psi_R {\cal D}A_{\mu, R}
e^{S } \ {\cal O} .
\eeqa
For simplicity, we first consider the $m_R=0$ case. In this case we have
\beqa
e^{S_{\rm eff}}&=&e^{S_0} \exp\left\{\sum_{n=1}^\infty
\left[
a^{2n} S_{2n}^0 + a^{2n-1} S_{2n-1}^1 \right] \right\}
\equiv  e^{S_0}\sum_{n=0}^\infty   a^n S^{(n)}\label{DefExpSEff}
\eeqa
where we define
$a^n S^{(n)}$ to be the sum of the $a^n$ terms in eq.\ \pref 
{DefExpSEff}.
For example, the first few terms are given as
\beqa
S^{(0)} &=& 1, \quad
S^{(1)} = S_1^1 , \quad
S^{(2)} = S_2^0 + \frac{(S_1^1)^2}{2!} .
\eeqa
Under the $T_1$ transformation, they behave as
\beqa
T_1 :\  S^{(n)}&\rightarrow&  (-1)^n S^{(n)} .
\eeqa
By expanding both action and operator, we have
\beqa
\langle {\cal O}_{\rm eff, R, sub}^{tp,d} (\{x\} ) \rangle_{S_{\rm eff}}
&=& \langle \left[{\cal O}^{tp,d}_R(\{x\})+ \sum_{l =1}^\infty a^l
{\cal O}_{R;tpd}^{[t+l],d +l}
(\{x\})\right]
\sum_{n=0}^\infty a^n S^{(n)} \rangle_{S_0} \nonumber \\
&=& \langle {\cal O}^{tp,d}_R(\{x\}) \rangle_{S_0} +
\sum_{l,n =0, l+n\not=0}^\infty a^{l+n} \langle {\cal O}_
{R;tpd}^{[l+t],d+l}(\{x\}) S^{(n)} \rangle_{S_0} .
\label{eq:correlation}
\eeqa
Since terms with $t+l+n =$ odd in the above expansion vanish
from the maximal twist condition (\ref{eq:maximal}),
terms with $t+l+n= 2s$ remain as
\beqa
\langle {\cal O}_{\rm eff, R,sub}^{tp,d} (\{x\} ) \rangle_{S_{\rm eff}}
&=& \delta_{t,0} \langle {\cal O}^{tp,d}_R(\{x\}) \rangle_{S_0} +
\sum_{s=1}^\infty  a^{2s-t} \sum_
{l=0}^{2s-t}
\langle {\cal O}_{R;tpd}^{[l+t],d+l}(\{x\})  S^{(2s-t-l)} \rangle_
{S_0} \nonumber \\
&=& \delta_{t,0} \langle {\cal O}^{tp,d}_R(\{x\}) \rangle_{S_0} +
\sum_{s=1}^\infty  a^{2s-t}  F_{d}^{2s-t}(\{x\}, g_R^2,
\log(\Lambda a), \mu_R ;  \mu_R^2a^2)\nn\\
\eeqa
where we define 
\beqa
F_{d}^{2s-t}(\{x\}, g_R^2, \log(\Lambda a), \mu_R ;
\mu_R^2a^2) &=&
\sum_{l=0}^{2s-t}
\langle {\cal O}_{R;tpd}^{[l+t],d+l}(\{x\}) S^{(2s-t-l)} \rangle_
{S_0} ,
\eeqa
which is an analytic function for small $\mu_R^2 a^2$ (the  last  
argument).
This expression tells us that
\beqa
\langle {\cal O}^{tp,d}_{\rm eff, R,sub} (\{x\} ) \rangle_{S_{\rm eff}}
&=& \left\{
\begin{array}{ll}
   \langle {\cal O}^{tp,d}_R(\{x\}) \rangle_{S_0} + a^2 F_d^2 + a^4
F_d^4 + \cdots, &  t=0 \\
aF_d^1 + a^3 F_d^3+\cdots, & t=1 \\
\end{array}\right. .
\label{eq:main_res}
\eeqa
This proves automatic O$(a)$ improvement at maximal twist that
scaling violations of $T_1$ invariant quantities are even functions
of $a$:  $a^{2n+1}$ contributions are completely
absent, while $T_1$ non-invariant quantities have only contributions
odd in $a$ and vanish in the continuum limit.
This completes our proof of automatic O$(a)$ improvement at
maximal twist.
(Here O$(a^n)$ ($n\ge 1$) represents contributions of the form
$a^{n+s} [\log(\Lambda a)]^k$ with $s,k=0,1,2,\cdots$).

Notice that this proof for automatic O$(a)$ improvement is not  
restricted to on-shell quantities, and
the equation of motion is not required at all for the proof.
It is also noted that the proof does not require
$\mu_R =0$:
Automatic O$(a)$ improvement is realized also for the massive theory.

If $m_R=O(a)$ (weak invariance), the proof goes through with just a  
little
modification
(see appendix \ref{appendixB}).
In the special case that $m_R$ is odd in $a$ ($m_R = a f(a^2)$), we
obtain
eq.(\ref{eq:main_res}) with a little modification in $F_d^n$, while
in more general cases with $m_R={\rm O}(a)$ the result becomes
\beqa
\langle {\cal O}_{\rm eff, R,sub}^{tp,d} (\{x\} ) \rangle_{S_{\rm eff}}
&=&\left\{
\begin{array}{ll}
   \langle {\cal O}_{R}^{tp,d}(\{x\}) \rangle_{S_0} + {\rm O}(a^2), &  
t=0 \\
O(a), & t=1 \\
\end{array}\right. .
\eeqa

%
\subsection{Ambiguity of the maximal twist condition in the lattice
theory}
%
\label{subsec:condition}
In this subsection we consider the maximal twist condition in the
lattice theory
and discuss the possible ambiguities of it.

In the Symanzik theory, maximal twist is uniquely defined by
the condition that an arbitrary $T_1$ non-invariant operator ${\cal O}^
{t=1\, p, d}$ has a vanishing expectation value,
\beqa\label{CondSym}
\langle
{\cal O}^{1p, d} \rangle_{S_0} &=& 0.
\eeqa
Provided this condition is fulfilled, the expectation values of all
$T_1$-odd operators vanish. Hence the particular choice for ${\cal O}^
{1p, d}$ is irrelevant, and in that sense the condition \pref
{CondSym} is unique.
In the lattice theory, however,  the maximal twist condition, which
may be
defined by
\beqa\label{CondLatt}
\langle  {\cal O}_{\rm lat, R,sub}^{1p,d} \rangle &=& 0,
\eeqa
depends on the choice of the operator ${\cal O}_{\rm lat,R,sub}^{1p,d}
$, and is therefore not unique.
In order to discuss this we make the Symanzik expansion of \pref
{CondLatt}, which gives
(see appendix \ref{appendixB} for unexplained notation and the  
derivation)
\beqa
0&=&
\langle {\cal O}_{\rm eff, R,sub}^{1p,d}\rangle
=
a H_d^0(\mu_R; a^2, m_R^2, m_R a, \mu_R^2 a^2) +
m_R H_d^1(\mu_R; a^2, m_R^2, m_R a, \mu_R^2 a^2).
\label{eq:maximal_mass}
\eeqa
The solution $m_R^{\rm maximal}$ to eq.\ (\ref{eq:maximal_mass}),
provided it is unique\footnote{
This seems plausible at small enough $a$ and $\mu_R$, since the
solution is unique in the Symanzik
theory.},
is of the form $m_R^{\rm maximal} = a f(a^2)$, due to the symmetry of
eq.\ (\ref{eq:maximal_mass})
under the transformation $m_R\rightarrow -m_R$ and $a\rightarrow -a$.
Therefore, according to the analysis in the previous subsection, scaling
violations in $T_1$-invariant quantities are even in $a$.

A different choice for the lattice operator in \pref{CondLatt} leads
to a different solution $\tilde{m}_R^{\rm maximal}   =a \tilde f(a^2)$,
so that the difference between the two definitions is of O$(a)$:
$\Delta m_R^{\rm maximal} = a \{f(a^2)-\tilde f(a^2)\}$.
Note that a solution $m_R^{\rm maximal}$ in  general depends on $\mu_R
$, inherited from the $\mu_{R}$ dependence of $H_d^\delta$.

Let us consider some examples for maximal twist in the lattice theory.
A simple one is given by
\beqa\label{condensate}
\langle \left(\bar\psi\psi\right)_{\rm lat, R,sub}\rangle &=& 0 .
\eeqa
Unfortunately,  this definition is not very useful in practice, since
the subtraction of power divergences necessary for $\langle\bar\psi
\psi\rangle$
prevents a reliable determination of this VEV in the lattice theory.
Instead one may take ${\cal O}_{\rm lat}(x,y) = A_\mu^a (x) P^a (y)$ or
${\cal O}_{\rm lat}(x,y) = \partial_\mu A_\mu^a (x) P^a (y)$ ($a=1,2$),
as was done in  Refs.\ \cite{Bietenholz:2004wv,Abdel-Rehim:2005gz,Jansen:2005mf}:
\beqa\label{LattCond2}
\langle A_\mu^a (x) P^a (y)\rangle &=& 0 \quad \mbox{ or } \quad
\langle \partial_{\mu}  A_\mu^a(x) P^a (y) \rangle = 0 ,
\label{eq:parity_con}
\eeqa
where $A_\mu^a$ and $P^a$ denote the axial vector current and pseudo
scalar density, respectively.
Yet another choice is \cite{Sharpe:2004ny}
\bea
\langle A_\mu^3 (x) P^3 (y)\rangle &=& 0. 
\eea
Depending on the choice for the axial vector current, either the
local or the conserved one,
the conditions \pref{LattCond2} lead to a different definition of
maximal twist. However, the difference will be again of O$(a)$.

We close this subsection with a final comment.
Any maximal twist condition in the lattice theory determines
a value for the bare mass $m_0$ as a function of the bare twisted
mass $\mu_0$. It has been suggested to tune the bare mass to its
critical value $m_0=m_{\rm cr}$ where the pion mass vanishes in the
untwisted theory. However, this condition is not related to  $T_1$
invariance. For example,
contributions from excited states violate eq.\ (\ref{eq:parity_con})
even at $m_\pi =0$.
Consequently, the pion mass definition does not correspond
to 
maximal twist according to the  $T_{1}$ invariance condition. 
%
\section{WChPT analysis for O$(a)$ improvement in tmQCD}
\label{sec:wcpt}
%
According to our analysis in the Symanzik effective theory, maximal
twist is determined by
requiring $T_1$ invariance of expectation values. For example, for
maximal twist we could impose
\beqa
\vev{\bar\psi \psi} &=& 0, \qquad
\vev{\bar\psi  i\gamma_5\tau_3\psi} = v(M_R)\not= 0.
\label{eq:vev}
\eeqa
In this section we study  this condition in Wilson Chiral
Perturbation Theory (WChPT)
\cite{Sharpe:1998xm,Rupak:2002sm,Bar:2003xq,Aoki:2003yv}, and check  
explicitly whether O$(a)$ improvement  is indeed realized. We also  
compare to some other definitions of
maximal twist, which have already been used in numerical simulations.

Automatic O($a$) improvement has been studied before in WChPT for
various definitions of the twist angle and also for different power  
countings,
which are determined by the relative size between the quark masses  
and the lattice spacing
\cite{Munster:2003ba,Aoki:2004ta,Sharpe:2004ny,Sharpe:2005rq}.
Our analysis follows closely the one in Ref.\ \cite{Aoki:2004ta}. We
work mainly in the regime where $m$ and $\mu$ are of O($a^
{2})$ unless stated otherwise.\footnote{Since the parameters $m$ and $
\mu$ in WChPT are renormalized parameters we
drop the subscript ``{\em R}'' in this section.} It is in this regime
where the phase structure of the theory
is determined by  the competition between the mass term and lattice  
spacing artifacts
\cite{Munster:2004am,Sharpe:2004ps}, and where the differences of the
various maximal twist definitions start to become relevant 
\cite{Aoki:2004ta}.
In contrast to Ref.\ \cite{Aoki:2004ta} we work at higher order and
include the terms of O($ma,\mu a,a^{3}$) in our analysis. These terms,
which were also included in Refs.\ \cite{Aoki:2005ii,Sharpe:2005rq},
provide a nontrivial check for automatic O($a$) improvement, since
they are odd in the lattice spacing and, according to our Symanzik
analysis, should not contribute to observables.

%
\subsection{Chiral Lagrangian and power counting}
\label{subsec:wcpt_tm}
%
In terms of the $SU(2)$ matrix-valued field $\Sigma$, which
transforms under chiral transformations as $\Sigma\rightarrow L\Sigma
R^{\dagger}$,  the chiral effective Lagrangian reads
\begin{align} \label{ChiralLag}
\mathcal{L}_\chi &=
   \frac{f^2}{4} \langle\partial_\mu \Sigma \partial_\mu \Sigma^\dagger
\rangle
-\frac{f^2}{4} \langle\hat{m}^{\dagger} \Sigma + \Sigma^\dagger\hat{m}
\rangle
-\frac{f^2}{4} \langle\hat{a}^{\dagger} \Sigma +
                 \Sigma^\dagger\hat{a}\rangle \notag \\
&\quad
+ (W_4 + W_5/2)\langle\partial_\mu \Sigma^\dagger \partial_\mu \Sigma
\rangle
        \langle\hat{a}^{\dagger} \Sigma + \Sigma^{\dagger}\hat{a}
\rangle\notag \\ &\quad
- (W_6 + W_8/2)\langle\hat{m}^{\dagger} \Sigma + \Sigma^\dagger\hat{m}
\rangle
        \langle\hat{a}^{\dagger} \Sigma + \Sigma^{\dagger}\hat{a}\rangle
        \notag \\
&\quad
- (W_6'+W_8'/2) \langle\hat{a}^{\dagger} \Sigma +
         \Sigma^{\dagger}\hat{a}\rangle^2\nn\\
         &\quad 
- W_{\rm c1} \langle\hat{a}^{\dagger} \Sigma +
         \Sigma^{\dagger}\hat{a}\rangle^3 
- W_{\rm c2}  \langle\hat{a}^
{\dagger}\hat{a}\rangle\langle\hat{a}^{\dagger} \Sigma +
         \Sigma^{\dagger}\hat{a}\rangle
- W_X \langle \hat{a}^\dagger \hat{m} + \hat{m}^\dagger\hat{a}\rangle.
\end{align}
The terms through O($a^{2}$) have been previously constructed \cite
{Sharpe:2004ps,Sharpe:2004bv}. Two terms of O($a^{3}$) in the last line,
which were also included in Refs.\ \cite{Aoki:2005ii,Sharpe:2005rq}
are easily derived with the spurion fields in
Ref.\ \cite{Bar:2003mh}. Note that the O($ m a $) term proportional to $W_X $ does not depend on
$\Sigma$. This term is usually neglected since it does not contribute to the pseudo scalar masses and decay constant. Here, however, we will need it since it gives a contribution to the condensates.
The coefficients $f, B $ are familiar  
low energy
coefficients of continuum chiral perturbation theory
\cite{Gasser:1983yg,Gasser:1984gg}, while all the $W$'s 
are  additional low-energy parameters associated with the nonzero  
lattice
spacing contributions \cite{Rupak:2002sm,Bar:2003mh}. As usual,
angled brackets denote traces over the flavor indices and the short- 
hand notation
\bea\label{DefM}
\hat{m} =
2B(m + i  \mu\tau_3)\,\equiv\,2 B m^{\prime}e^{i\omega_{L}\tau_{3}},
\qquad\hat{a }= 2 W_0 \, a \,,
\eea
is used \cite{Bar:2002nr}.
The mass parameters $m$ and $\mu$ denote the renormalized untwisted
and twisted mass \cite{Frezzotti:2000nk}, which are defined according
to\footnote{The renormalization constants $Z_{m},Z_{\mu}$ are
related to the renormalization constant $Z_{A}$ of the axial vector,
$Z_{A} = Z_{m}/Z_{\mu}$,
which follows from the vector and axial vector Ward identities \cite 
{Frezzotti:2000nk}.}
\bea\label{Def:RenMasses}
m& =& Z_{m}(m_{0} - m_{\rm cr}),\qquad \mu\,=\, Z_{\mu} \mu_{0}\,.
\eea
Even though the critical mass $m_{\rm cr}$ includes the additive
shift proportional to $1/a$, it does not include certain
contributions coming from the O($a,a^{2}$) terms in the chiral
Lagrangian \cite{Sharpe:1998xm}. For example, the third term in the
first line of \pref{ChiralLag} gives rise to an O($a$) shift in the
critical mass.\footnote{This term is often absorbed in the untwisted
mass, giving rise to the so-called shifted mass 
\cite{Sharpe:1998xm,Sharpe:2004ny,Sharpe:2005rq}.}

Our power counting is based on the assumption that $m \approx a^{2}$
\cite{Aoki:2003yv, Aoki:2004ta,Aoki:2005mb}, where $m$ stands for
both the untwisted and the twisted mass and for $p^{2}$ (proper
powers of $\Lambda_{\rm QCD}$ are, as usual in this type of argument,
understood). Since $m$ and $a$ are smaller than one we have the
inequalities
\be\label{AssumptionPowerCounting}
m \approx a^{2} > m a \approx a^{3} > m^{2} \approx m a^{2}\approx a^
{4}.
\ee
According to this power counting the terms of O($m,a^{2}$) in the
chiral Lagrangian are of leading order (LO), while the O($m a,a^{3})$
contributions are of next to leading order (NLO).
Note that the size of the O$(a)$ term does not matter for the power
counting, since it only contributes to the critical quark mass.
%
\subsection{Gap equation}
%
Starting from the chiral Lagrangian a gap equation for the ground
state of the chiral effective theory can be derived. From the NLO
expression of the chiral Lagrangian we find the potential
\beqa\label{PotentialEnergy2} 
V_{\chi} &=&  \frac{f^2}{4} 2Bm^\prime \langle P^{\dagger} \Sigma + \Sigma^
\dagger P\rangle
+\frac{f^2 }{4} 2W_{0}a(1 + \tilde{c}_{3}a^{2})\langle\Sigma +
                 \Sigma^\dagger\rangle - \frac{f^2}{16}c_{2} a^{2}
\langle \Sigma +
         \Sigma^{\dagger}\rangle^2\nn\\
                 &+& \frac{f^2}{16}\tilde{c}_{2} a m^\prime \langle \Sigma +
         \Sigma^{\dagger}\rangle\langle P^{\dagger} \Sigma + \Sigma^
\dagger P\rangle
     + \frac{f^2}{64} c_{3} a^{3} \langle \Sigma +
         \Sigma^{\dagger}\rangle^3
+\frac{2Bf^2m^\prime}{4}c_X a\langle P^\dagger + P \rangle .
\eeqa
Here we introduced $P  =\exp{i\omega_{L}\tau_{3}}$ with $\tan \omega_
{L} = \mu/m$, and the following combinations of low energy parameters:
\bea
c_{2} & = & -32\, (2W^{\prime}_{6}+W_{8}^{\prime}) \frac{W_{0}^{2}}{f^
{2}},\qquad
\tilde{c_{2}} \, = \, 32\, (2W_{6}+W_{8}) \frac{W_{0}B}{f^{2}},\nn \\
c_{3} & =&64W_{\rm c1} \frac{(2W_{0})^{3}}{f^{2}},\qquad\qquad\quad
\tilde{c_{3}} \, = \, 32W_{\rm c2}\frac{W_{0}^{2}}{f^{2}},
\qquad
c_X = 8W_{X }\frac{W_{0}}{f^{2}} 
\label{Def:c2}.
\eea
These parameters are dimensionful and have $[c_{2}]=4$, $[\tilde{c}_
{2}]=[\tilde{c}_{3}]=2$, $[c_{\rm 3}]  =5$ and $[c_X]= 1$.

Since a twisted mass term breaks flavor symmetry we make the ansatz
\beqa\label{AnsatzVEV}
\Sigma_0 &=& e^{i \phi \tau_3}
\eeqa
for the ground state, and
this ground state is determined by $d V_{\chi}/{d\phi}=0$ 
with
\begin{eqnarray}
V_{\chi}
&=&  2Bf^2( mt +\mu\sqrt{1-t^2})+2f^2 W_0 a(1+\tilde c_3a^2)t
-f^2 c_2 a^2 t^2 \nn \\ &+&f^2\tilde c_2 a ( mt +\mu\sqrt{1-t^2}) t
+ f^2 c_3a^3t^3 + 2Bf^2 c_X ma .
\label{PotentialEnergy3}
\end{eqnarray}
Taking the derivative with respect to $\phi$ in
\pref{PotentialEnergy3} we obtain a {\em gap equation} for
\bea
t& =& \cos\phi,
\eea
which can be brought into the form
\beqa
\sqrt{1-t^2}\left[\chi - t +2\beta_m t +\gamma t^2\right] &=&
\alpha \left[ t - \beta_\mu ( 1- 2t^2)\right],
\label{eq:gap}
\eeqa
where we introduced the dimensionless parameters
\beqa
\alpha &=& \frac{2B\mu}{2 c_2a^{2}},\qquad
\chi \,=\, \frac{ 2 B m + 2 W_{0}a(1 + \tilde{c}_{3}a^{2}) }{2c_2a^ 
{2}},\nn \\
\beta_{m}& =& \frac{ \tilde{c}_{2}am  }{2c_2a^{2}}, \qquad
\betamu\, = \, \frac{\tilde{c}_{2}a}{2B},\qquad
\gamma \, =\, \frac{3c_{3}a^{3}}{2c_{2}a^{2}}.\label{Defbeta}
\eeqa
In the following we will assume
\bea
|\betamu|& <&1,\qquad |\betam|\, <\,1,\qquad |\gamma| \, < \,1,
\label{Reqbeta}
\eea
which can be justified by a naive dimensional analysis
when all dimensionfull constants are
assumed to be of O($\Lambda_{{\rm QCD}})$ together with the
conditions $a\Lambda_{{\rm QCD}}<1$ and  $m/\Lambda_{{\rm QCD}}<1$.

Note the sign convention for the coefficient $c_{2}$. A positive sign
corresponds to the scenario with spontaneous parity-flavor breaking
\cite{Sharpe:1998xm}, which guarantees the existence a massless pion
\cite{Aoki:1983qi}. A negative coefficient $c_{2}$ results in a
scenario  with a first order phase transition
\cite{Munster:2004am,Sharpe:2004ps}.\footnote{See also Ref.\ \cite 
{Sharpe:2005rq}
where it has been shown that the NLO terms in the chiral Lagrangian do
not change the existence of two qualitatively different scenarios for
the phase diagram.}
The details of the discussion of O($a$) improvement differ depending on
the scenario for the phase diagram.
In the rest of this section we are mainly interested in the scenario  
with $c_{2}>0$, where spontaneous parity-flavor breaking causes some  
subtleties for automatic O($a$) improvement.
These subtleties are absent for $c_{2}<0$ and we come back to this  
scenario
at the end of this section.
 %
\subsection{Condition for O$(a)$ improvement in WChPT}
\label{subsec:wcpt_condition}
%
Taking derivatives of $V_\chi$ with respect to $m$ and $\mu$,
the two VEVs $\vev{\bar\psi \psi} $ 
and $\vev{\bar\psi i\gamma_5\tau_3 \psi}$ are easily computed with
the result
\beqa 
\langle \bar\psi \psi  \rangle &\equiv& 
\frac{d V_{\chi}}{d\, m}
=  2 f^2 B\left[(1+\beta_\mu t) t +c_X a\right]\nn\\
\langle \bar\psi i\gamma_5\tau_3 \psi  \rangle &\equiv& 
\frac{d V_{\chi}}{d\, \mu}
=  2 f^2 B(1+\beta_\mu t) \sqrt{1-t^2}.
\eeqa
Therefore, the $T_{1}$ invariance condition (\ref{eq:vev}) corresponds to $t = -c_X a + O(a^3)$ 
in WChPT.
If general scalar and pseudoscalar operators are employed for
$\bar\psi\psi$  and 
$\bar\psi i\gamma_5\tau^3\psi$, 
these results are modified by O($a$) stemming from the effective operators in the Symanzik effective theory \cite{ABT}. This leads to
\beqa
\langle \bar\psi \psi  \rangle 
&=& \frac{2 f^2 B}{Z_S}\left[(1+c_S\, a\, t) t +\tilde c_S\, a\right],\nn\\
\langle \bar\psi i\gamma_5\tau_3 \psi  \rangle
&=&\frac{2 f^2 B}{Z_P}(1+c_P\, a\, t) \sqrt{1-t^2}. 
\eeqa
Nevertheless, even in this case the condition (\ref{eq:vev}) leads to a similar result:
$t=-\tilde c_S a +O(a^3)$.

We would find a similar result using the alternative condition $\vev{A_
\mu^2 P^2}=0$, which is equivalent to
$\cot \omega_{\rm WT} = 0$ with
\beqa\label{Def:cotw}
\cot \omega_{\rm WT} &\equiv& \frac{\vev{A_\mu^2 P^2}}{\vev{ V_\mu^1
P^2} }.
\eeqa
Note that $A_\mu^2$ is $T_1$-odd while $V_\mu^1$ and $P^2$ are $T_1$-
even. Provided 
Noether currents are used for $A_\mu^2$ and $V_\mu^1$ 
one finds (see also appendix \ref{appendixC})
\beqa
\langle 0\vert A_{\mu= 0}^2 \vert \pi_2\rangle &=& f m_\pi t
(1+c_0\, a\, t) ,\quad\qquad
\langle 0\vert V_{\mu= 0}^1 \vert \pi_2\rangle \,=\, f m_\pi \sqrt{1- 
t^2}\, (1+c_0\, a\, t),
\eeqa  
where we defined the coefficient
\bea
c_{0} & =&  16\, (2W_{4}+W_{5}) \frac{W_{0}}{f^{2}},
\eea
in analogy to the definitions in Eq.\ \pref{Def:c2}.
This leads to $\cot \omega_{\rm WT} =\cot \phi$, hence the condition $
\vev{A_\mu^2 P^2}=0$  implies $t=0$.
The result differs if one uses general non-Noether currents.
Additional contributions of O($a$) appear in the effective operators
in the Symanzik expansion, which carry over to the chiral effective
theory as well:
\beqa
\langle 0\vert A_{\mu= 0}^2 \vert \pi_2\rangle
&=& \frac{ f m_\pi}{Z_A} \Big[t(1 + c_A a t) - \tilde{c}_{A}a
\Big], \quad
\langle 0\vert V_{\mu= 0}^1 \vert \pi_2\rangle
\,=\, \frac{ f m_\pi}{Z_V} \sqrt{1-t^2}
\left[ 1 + c_V a t \right] .
\eeqa 
Note that here the currents  on the left hand side are bare
currents, as one can infer from the explicit appearance of the
renormalization constants $Z_A,Z_V$.  The way we have
written the
expectation values correspond to what can be directly measured in a
lattice simulation without the knowledge of $Z_A,Z_V$. For $\cot
\omega_{\rm WT}$ we find
\beqa\label{cotw:local}
\cot \omega_{\rm WT} 
&=& \frac{t(1+ c_A a t)-\tilde c_A a}{\sqrt{1-t^2}}\times \frac{Z_V}{Z_A}
\times \frac{1 } {1+ c_V a t },
\eeqa
and the maximal twist condition $\cot \omega_{\rm WT}=0$ gives
$t= \tilde c_A a + O(a^3)$, which has the same form as in the case
of the VEVs.
Note here that imposing a non-vanishing value for $\cot
\omega_{\rm WT}$ is sensitive to the ratio $Z_V/Z_A$ as well as to
$c_A$, $\tilde c_A$ and $c_{V}$.
As a final example we consider the condition $\vev{A_\mu^{3} P^3}=0$ introduced in \cite{Sharpe:2004ny}. Since
\bea
\vev {0| P^3| \pi^{3}} & = & \frac{ f B}{Z_P}\left[t-(1-2t^2)c_P\, a\right]
\eea
we again find $t = c_P a + O(a^3)$. 

To summarize, imposing $T_{1}$ invariance we find the condition $t = X a + O(a^3)$ with some constant $X$. This constant depends on the specific choice for the operator in the matrix element. Nevertheless, all definitions guarantee automatic O($a$) improvement, as we want to show next. 

It is instructive to first consider the simpler condition
$t=0$ (which is equivalent to a vacuum
angle $\phi = \pm \pi/2$).
In this case the pseudo scalar mass and decay constant of
charged pions, $m_\pi^2$ and $f_\pi$ are given by (see
also appendix \ref{appendixC})
\beqa
m_\pi^2 &=& \frac{2 B \mu}{\sqrt{1-t^2}}\frac{1 + \beta_\mu t }
{1 + c_0 a t }, \label{Mpi}\\
f_\pi &=& f\sqrt{1-t^2}
\left[1 + c_0 a t \right] \label{Fpi} .
\eeqa
These expressions are valid for arbitrary $t$, but for $t=0$ they
turn into the results familiar from leading order
continuum ChPT,
\bea
m_{\pi}^2 &=& 2 B \mu, \qquad\qquad
f_{\pi} \,=\, f .
\eea
Apparently there are no O$(a,a^{3})$ corrections in these
results.
In addition, the O$(a^2)$ corrections are also absent, but this is
not as surprising as one might first think. The charged pions are the
Goldstone bosons associated with the spontaneous breaking of
flavor and parity in the theory without a twisted mass term. Hence
they must become massless when one enters the broken phase, i.e. when
$\mu$ goes to zero. With the same argument one would also conclude
that no terms of order O$(a,a^{3})$ terms are present. The same
argument, however, does not apply to the O($\mu a$) terms, and their
absence is indeed a non-trivial demonstration of automatic O($a$)
improvement once $T_{1}$ invariance is imposed.

It is now simple to show that we can relax
$t=0$ 
to the weaker condition $t = {\rm
O}(a)$ without loosing automatic O($a$) improvement.\footnote{A
similar argument that the theory is O$(a)$ improved for $t={\rm O}(a) 
$ has
also been given independently by S.\ Sharpe \cite{SharpeNara}.}
Suppose that $t=X a$ with some constant $X$. If we insert this into
\pref{Mpi} and \pref{Fpi} we find (after expanding the denominator)
\bea
m_\pi^2 &=& 2 B \mu  \Big( 1 + \left[\frac{\tilde{c}_{2}}{2B} - c_{0}
+ \frac{X}{2}\right] X a^{2}\Big) + {\rm O}(\mu a^{4}),\label{mpiweak}\\
f_\pi &=& f\Big(1 + [c_{0} - X/2] X a^{2}\Big) + {\rm O}( a^{4}).\label{fpiweak}
\eea
Again no O$(a,\mu a)$ corrections appear. 
This demonstrates, within WChPT and at least for the two observables
we have chosen,
that $t = {\rm O}(a)$, which follows from imposing $T_{1}$ invariance,
 is a sufficient condition for automatic O$(a)$
improvement.
%
\subsection{Other conditions for maximal twist and O$(a)$ improvement}
\label{subsec:wcpt_3def}
%
In the following we want to compare the condition of $T_{1}$
invariance to some other conditions for maximal twist which are proposed
in the literature. In particular we are interested in definitions where 
the untwisted mass $m_0$ is set to a particular value and kept fixed as
one varies the twisted mass $\mu_0$. Such definitions obviously have a 
practical advantage for numerical simulations. Finding the $\mu_0$
dependent value $m_0$ such that a matrix element like the ones in 
\pref{condensate} or \pref{LattCond2} vanishes is computationally 
quite demanding, in particular in dynamical simulations. One can save 
a substantial amount of computer time if one does not need to do this
tuning for each twisted mass one wants to simulate, but rather stay at
one fixed value of $m_0$.
However, such definitions do violate $T_{1}$
invariance for most $\mu_0$ values and it is therefore not obvious how
this affects automatic 
O$(a)$ improvement. This is the issue we want to study in this section.

The following two definitions keep the untwisted mass constant and both 
have already been employed in quenched numerical 
simulations \cite{Bietenholz:2004wv,Abdel-Rehim:2005gz,Jansen:2005gf,
Jansen:2005kk}:
\begin{enumerate}
\item PCAC mass definition. For a given (bare) twisted mass $\mubare$ 
the untwisted mass $\mbare (\mubare)$ is first tuned such that the PCAC quark
      mass, defined by
\beqa\label{DefPCAC}
    2 m_{\rm PCAC} =\frac{ \vev{\partial_{\mu} A_\mu^2 P^2}}{\vev{P^2
P^2}},
\eeqa 
vanishes.  Then $\mbare (0) = \lim_{\mu_0\rightarrow 0}\mbare(\mubare)$ 
is used as the choice for $\mbare$ of the PCAC mass definition for 
all $\mubare$.
Therefore, this definition is independent of $\mubare$.\footnote{Note
that one can choose $\mbare(\mubare)$ at a fixed non-vanishing value
      $\mu_{0}$ as a (different) PCAC mass definition.
Employing such a definition requires the determination of
      $\mbare(\mubare)$ at only one $\mu_{0}$ value. Taking the 
$\mu_{0}\rightarrow 0$ limit, on the other hand, requires the 
determination of $\mbare(\mubare)$ for various twisted mass values and 
a subsequent extrapolation. Hence the latter is numerically more demanding.}
\item 
Pion mass definition. 
The bare untwisted quark mass is set to its critical quark mass where 
the pion mass vanishes at $\mubare =0$.
In practice this value is usually obtained in the untwisted theory by performing an extrapolation of $m_{\pi}^{2}$ data to the massless point.
\end{enumerate}
In order to study these two definitions we have to translate (''match'') them to the corresponding ones in WChPT:
\begin{enumerate}
\item PCAC definition: The denominator in Eq.\ \pref{DefPCAC}  is
found to be given by
\bea
\langle 0 \vert P^{2} \vert \pi_2 \rangle & = & \frac{B f}{Z_P} (1 +  
\beta_\mu t)
\eea
so that the PCAC condition reads
\beqa
m_{\rm PCAC}&=& \mu \frac{Z_P}{Z_A} \frac{t (1 +   
c_A\, a\, t) - \tilde{c}_{A}\, a }{\sqrt{1-t^2}(1 + c_{0}\, a\, t)} = 0.
\eeqa 
This leads to $t=\tilde{c}_{A}a +{\rm O}(a^{3})$ for any non-zero
      $\mu$. Keeping, for simplicity, only the leading term and setting
      $t= \tilde{c}_{A}a$ 
into the gap equation, one finds 
\bea\label{WChPTCondPCAC} 
\chi = \tilde{c}_{A}a [1 - 2\beta_{m} - \gamma\tilde{c}_{A}a ]
\eea
in WChPT.
In the following we will assume this $\chi$ to be smaller than one. This
      is in accordance with our previously made assumption that all
      dimensionful 
coefficients 
are of order $\Lambda_{\rm QCD}$ and $a\Lambda_{\rm QCD}< 1$.

\item Pion definition: We need the expression for the pion mass in
the untwisted theory. We cannot simply take the $\mu\rightarrow0$
limit of Eq.\ \pref{Mpi}, since $t$ is equal to 1 in this limit and  
the whole
expression is ill-defined. Instead, we first use the gap equation and
rewrite the pion mass as
\bea\label{Pionmass2}
   m_{\pi}^2 &=& \frac{2c_{2}a^{2}}{1+c_{0}at}\bigg[\frac{\chi}{t} -  
1 +2
\beta_{m} + \gamma t + 2\beta_{\mu}\alpha\frac{\sqrt{1-t^{2}}}{t}\bigg].
\eea
Here the limit $\mu\rightarrow0$ is well-defined and
the condition
$m_\pi^2 =0$ reads
\beqa\label{WChPTCondPion}
\chi &=& 1 - 2\beta_m - \gamma.
\eeqa
\end{enumerate}

In order to check whether O$(a)$ improvement is realized we have to
verify that  $t$ is at least of ${\rm O}(a)$. 
To do so we have to solve the gap equation (\ref
{eq:gap}) with the $\chi$ values in \pref{WChPTCondPCAC} and \pref
{WChPTCondPion}, respectively. It is not necessary to solve the gap  
equation exactly, approximate solutions will be sufficient for our  
purposes.

Let us assume $t\ll 1$, since we are interested in the small $t$ case.
In this case we can neglect the $t^2$ terms in \pref{eq:gap} and obtain the approximate solution
\bea\label{GESOLapprox}
t\simeq \frac{\alpha \betamu + \chi}{\alpha + 1 - 2\betam} .
\eea
For the PCAC mass condition we set $\chi$ to the value in \pref{WChPTCondPCAC}  and find
\beqa\label{eq:asympt_1}
t\simeq a \frac{\tilde c_2 \mu + 2\tilde{c}_{A} \left(c_2 +\tilde c_2 W_0/B
\right) a^{2}}{2 B\mu +2\left(c_2 +\tilde c_2 W_0/B
\right) a^2} .
\eeqa
Here we rewrote \pref{WChPTCondPCAC} as
\bea\label{am} 
am & =& -\frac{W_{0}}{B}a^{2} + {\rm O}(a^{4}), 
\eea
and dropped all but the leading term proportional to  $a^{2}$.
Taking into account that the denominator in eq.\ \pref{eq:asympt_1} is
always of O$(a^2)$ or 
larger for $c_2+\tilde c_2 W_0/B > 0$, together with our convention
that $B > 0$ and $\mu>0 $, $t$ is always of O$(a)$ and
our assumption that $t\ll 1$ is consistently satisfied.\footnote{Recall that we here consider the case with $c_{2}>0$.
If $c_2+\tilde c_2 W_0/B < 0$, the
assumption $t\ll 1$ could be
violated at some value of $\mu$. Therefore, in the latter case, we
exclude such values of $\mu$ in the following consideration.}
The solution in eq.\ \pref{eq:asympt_1}  is of the form $t=aX$ which we used at the end of the 
previous section (cf.\ eqs. \pref{mpiweak} and \pref{fpiweak}), with $X$
representing the fraction in \pref{eq:asympt_1}. However,
here the value of $X$ does depend on 
the relative size between $a$ and $\mu$. 
For small $a$ and fixed $\mu$ such that $2 B\mu \gg 2\left(c_2 +\tilde c_2  W_0/B\right) a^2$, we can expand the denominator and find
\bea\label{eq:asympt_2}
t&\approx &a\frac{\tilde{c}_{2}}{2B} ,
\eea
so $X = \tilde{c}_{2}/{2B}$ and $t={\rm O}(a)$. Hence our discussion in the last section can be applied and we find O($a^{2}$) scaling violations in physical observables.
On the other hand, for larger $a$ such that 
$\mu = {\rm O}(a^{2})$ we expect a modification of the  
simple linear $a$ dependence of $t$, and this leads to distortions of
the expected O($a^{2}$) scaling violations. 
In the extreme case of small fixed $\mu$ and large $a$ such that 
 $2 B\mu \ll 2\left(c_2 +\tilde c_2 W_0/B  
\right) a^2$ we find
\bea\label{eq:asympt_3}
t&\approx & a\tilde{c}_{A},
\eea
as expected from the definition according to the PCAC definition. Even though we recover a constant $X$, it is different from \pref{eq:asympt_2}. Since the sign of the low energy constants are 
a priori not known, it is even possible that the slope of $t$ changes sign, depending on the size of $a$. 
This is of potential danger when one analyzes numerical data assuming a simple O$(a^{2})$ scaling violation. The non-trivial $a$ dependence of the r.h.s.\ in eq.\ \pref{eq:asympt_1} is likely to obscure automatic O($a$) improvement in the region where $\mu$ is of ${\rm O}(a^{2})$.\footnote{We emphasize that the non-trivial behaviour in $t = aX(a^{2})$ does not mean that there are terms linear in $a$ in physical observables.}

Let us now turn to the pion mass definition.
Inserting \pref{WChPTCondPion} into the approximate solution \pref 
{GESOLapprox} we obtain
\bea\label{eq:asympt_4}
t\simeq a \frac{\tilde c_2 \mu +2(c_2 +\tilde c_2 W_0/B)a }{2 B 
\mu +2\left(c_2 +\tilde c_2 W_0/B
\right) a^2} .
\eea
In order to derive this result we rewrote \pref{WChPTCondPion} as in eq.\ \pref{am}
and used, for simplicity,  only the leading term proportional to  $a^ 
{2}$.
As before we find $t\approx a\tilde{c}_{2}/2B = {\rm O}(a)$ for small  
values of $a$, and, since the denominator is the same as for the  
PCAC definition we expect the modifications to become visible once the lattice spacing is such that
$\mu$ is of  ${\rm O}(a^{2})$. The details of the modification will  
be different because  the numerator differs compared to the PCAC mass definition.

However, the crucial difference between the PCAC and the pion mass definition is that
the approximation \pref{GESOLapprox} will eventually break  
down for the pion mass definition, since $t$ goes to 1 for a  
vanishing $\mu$. In that case the $t^{2}$ terms can no longer be ignored.
Interestingly, the approximate solution \pref{eq:asympt_4} gives the
correct value $t=1$ at $\mu=0$ even though this approximation can 
not be justified. A more careful analysis finds
\cite{Aoki:2004ta}
\beqa\label{tPiondef}
t & \simeq & 1 -\delta, \quad
\delta = \frac{1}{2}\left[\frac{\mu ( 2 B +\tilde c_2 a)}{(c_2 +
\tilde c_2
W_0/B)a^2}\right]^{2/3} = {\rm O}\left( \left[\frac{\mu}{a^2}\right]^{2/3}
\right).
\eeqa
Therefore, the condition $t= {\rm O}(a)$ for automatic O$(a)$  
improvement
is satisfied only for small lattice spacings where  $2 B \mu \gg 2\left(c_2 +\tilde c_2 W_0/B  
\right) a^2$.
Even though this bound is asymptotically satisfied
as $a\rightarrow 0$, at a given non-vanishing lattice spacing the  
scaling violation becomes
sizable for small twisted quark masses.
In particular in the region $2 B \mu \ll 2(c_2+ 
\tilde c_2 W_0/B) a^2$, where $t\rightarrow 1$ in  
the $\mu \rightarrow 0$
limit, automatic O$(a)$ improvement fails.

This failure is seen when one inserts \pref{tPiondef} into expression \pref{Fpi} for the decay constant, for example. Ignoring the small correction coming $\delta^{2}$ and higher powers we find
\bea
f_\pi &=& f\sqrt{2\delta}(1 + c_0 a - c_0 a\delta),
\eea 
which obviously has a term linear in $a$.
We emphasize that the reason for the presence of terms linear in $a$ is the leading 1 in $t = 1-\delta$, and not the correction $\delta$ with the peculiar dependence on fractional powers of $a$, even though the overall factor $\sqrt{2\delta}$ complicates the whole $a$ dependence. The leading constant term is the crucial difference to the PCAC mass definition, where $t={\rm O}(a)$, and this constant term spoils automatic O$(a)$ improvement for lattice spacings where $2 B \mu \ll 2(c_2+ \tilde c_2 W_0/B) a^2$.

We summarize the results in this section as follows.
Although the PCAC mass and pion mass definitions lead to
O$(a)$  improvement for small enough lattice spacings,
the asymptotic O$(a^{2})$ behaviour can only be seen at lattice spacings
where the bound $\mu \gg a^{2}\Lambda_{\rm QCD}^{3}$ is realized for a given $\mu$.
If this bound is not satisfied,
the naively expected scaling violation is compromised, in particular  
for the pion mass definition, but also for the PCAC mass definition, even though to  a much lesser extent.
Note that the bound $2 B \mu \gg 2(c_2+\tilde c_2 W_0/B) a^2$
excludes automatic O$(a)$ improvement for the massless theory 
in the case of the pion mass definition,  
which is not unexpected since it is identical to the massless  
(untwisted) Wilson theory.
%
\subsection{ Automatic O$(a)$ improvement for the $c_2<0$ case}
\label{subsec:wcpt_negative}
%
Since the condition for automatic O$(a)$ improvement discussed in the
previous section
does not depend on the details of  the lattice QCD dynamics,
it seems applicable quite generally.
However, there are circumstances when conditions like $\vev{ A_\mu^ 
{2} P^2 } = 0$ or $\cot \omega_{\rm WT}=0$ cannot be satisfied. This  
is the case when the 1st order phase transition scenario of Refs.\   
\cite{Sharpe:1998xm,Farchioni:2004us} is realized.

Let us consider this case in WChPT. For simplicity we work at LO only  
and set $\tilde c_2 = c_3 =0$  in the following argument. In this  
case, if $c_2 < 0$,
a first order phase transition appears at $\chi=0$ 
\cite{Sharpe:1998xm,Munster:2004am} and $t$ is given by
\cite{Aoki:2004ta}
\beqa
t &=&\left\{
\begin{array}{cll}
\phantom{-}\sqrt{1-\alpha^2}, & \alpha^2 < 1, &  \chi\rightarrow 0^+ \\
-\sqrt{1-\alpha^2}, & \alpha^2 < 1, &  \chi\rightarrow 0^- \\
0, & \alpha^2 \ge 1, & \chi\rightarrow 0 \\
\end{array}\, ,
\right. \quad \alpha=\frac{\mu}{\mu_{\rm min.}}  . 
\eeqa
Although the condition $t=0$
can be realized for $\chi = 0$, the twisted mass
$\mu$ must satisfy the bound $\mu^2 \ge \mu_{\rm min.}^2$, where
\beqa
\mu_{\rm min.}^2 &=& \left(\frac{2 c_2 a^2}{2 B}\right)^2  .
\eeqa
Therefore, automatic O$(a)$ improvement can only be realized
for $\mu^2 \ge \mu_{\rm min.}^2$,
in contrast to the parity conservation definition
for the $c_2 > 0$ case, where no restriction on $\mu$ needs to be  
imposed.
Note, however, that the same restriction  on $\mu$ (at LO) is  
required for the pion mass definition in the
$c_2 > 0$ case.

This argument does not change qualitatively when one includes the NLO  
terms, as has been done in Ref.\ \cite{Sharpe:2005rq}. The phase  
transition line is no longer a straight line in the $m_{0}-\mu_{0}$  
parameter plane. If the term with $\tilde c_2$ is included the  
maximal twist condition which gives
$t=0$ becomes $\mu$ dependent and reads $\chi = -\mu \tilde c_2 a/ 
(2c_2 a^2)$. Nevertheless, the conclusion that one has to stay above  
the phase transition line in order to be able to satisfy the maximal  
twist condition remains unchanged.

%
\section{Conclusion}
\label{sec:conclusion}
%
In this paper we gave
an alternative proof  for automatic O$(a)$ improvement in twisted
mass lattice QCD at maximal twist.
Whereas previous proofs \cite{Frezzotti:2003ni,Frezzotti:2005gi} used
symmetries of the bare lattice theory such as $m_q\rightarrow -m_q$ and
$r\rightarrow -r$,
we have used only symmetries of the
leading part of the Symanzik effective theory in our proof.
A more important observation, however, is that
a precise definition for the twist angle, and therefore a condition
for maximal twist,
is determined dynamically by the ratio of two vacuum expectation  
values in the Symanzik theory:
\beqa
\cot \theta &=&\frac{\langle \bar\psi\psi\rangle}{\langle \bar\psi i
\gamma_5
\tau^3 \psi\rangle}\,.
\eeqa
At $\theta = \pm\pi/2$, which is equivalent to $T_1$ invariance
of the vacuum in the continuum theory, scaling violations for
all quantities are shown to be even powers in $a$,
as long as they are invariant under the $T_1$ transformation.
Non-invariant quantities, on the other hand, vanish as odd powers in  
$a$.
It is also shown that the ambiguity for the maximal twist condition in
the lattice theory does not spoil
automatic O$(a)$ improvement.

We also studied the  $T_1$ invariance condition in WChPT. As expected, for the pseudo scalar mass and the decay constant we find automatic O($a$) improvement.

We finally compared the  $T_1$ invariance condition to two other definitions for maximal twist, the  PCAC mass and the pion mass definition. Both definitions have already been used in numerical simulations. These definitions have the practical advantage that the untwisted mass is tuned to a fixed value independent of the twisted mass $\mu$.
Although both definitions give asymptotic $a^2$ scaling  
violations for $\mu \gg a^{2}\Lambda^{3}_{\rm QCD}$, we have  
shown that the expected $a^2$ scaling can be obscured once this bound is violated.
Hence naive continuum extrapolations  
can be deceiving and may lead to wrong results
for these definitions of maximal twist.
Here a WChPT analysis seems to be indispensable for a controlled continuum  
extrapolation.

%
\section*{Acknowledgments}
%

We acknowledge useful discussions with R.~Frezzotti, M.~Golterman, K.~Jansen, S.~Sharpe, S.~Sint and P.~Weisz. We would like to thank M.~Golterman and S.~Sharpe for a careful reading of the manuscript.  We also thank the Institute for Nuclear Theory at the University of  
Washington for its hospitality and the Department of Energy for  
partial support during the completion of this work.

This work is supported in part by the Grants-in-Aid for
Scientific Research from the Ministry of Education,
Culture, Sports, Science and Technology.
(Nos. 13135204, 15204015, 15540251, 16028201),
O. B.\ is supported in part by the University of Tsukuba Research
Project.

%
\begin{appendix}
%

\section{Some details for the proof of O($a$) improvement}
\label{appendixB}
\subsection{Derivation of Symanzik action}
We apply the Symanzik expansion to all operators which appear in the  
lattice
action, as
was done to ${\cal O}^{01,3}_{\rm lat}$ in the main text,
After a little algebra we obtain the following expression for the
effective action:
\beqa
S_{\rm tmQCD} \Leftrightarrow S_{\rm eff} &=& S_0  + m_q S_m +\sum_
{n=1}^\infty\left[
a^{2n}S_{2n}^0 + a^{2n-1} S_{2n-1}^1\right],
\eeqa
where
\beqa
S_0 &=& \int d^4 x\ \left[ Z_F {\cal O}^{00,4}_A(x)  + Z_F Z_\mu \mu\
{\cal O}^{01,3}(x) +
Z_G  {\cal O}_B^{00,4}\right],\\
S_m &=& \int d^4 x\ Z_F\cdot Z_m {\cal O}^{10,3}(x),
\eeqa
and
\beqa
S_{2n}^0 &=& \int d^4x\ \sum_i C_{00,2n+4}^i \ {\cal O}^{00,2n+4}_i(x)
+\sum_i C_{01,2n+3}^i\ \mu
\ {\cal O}^{01,2n+3}_i(x),\\
S_{2n-1}^1 &=& \int d^4x\ \sum_i  C_{10,2n+3}^i\ {\cal O}^{10,2n+3}_i
(x)+\sum_i C_{11,2n+2}^i\
\mu \ {\cal O}^{11,2n+2}_i(x)  .
\eeqa
In the definitions of $S_n^{t_n}$,  the superscripts $t_n=0,1$
represent the transformation property
under $T_1$:
\beqa
T_1:\  S_n^{t_n}&\rightarrow& (-1)^{t_n} S_n^{t_n}.
\eeqa
All coefficients which appear in the expressions above, such as $Z_{\{F,
\mu,m,G\}}$ and $C_{tp,d}^i$,
are dimensionless functions of $g^2$, $\log(\Lambda a)$,  $m_q a$
and $\mu_0^2a^2$.
They are given in terms of  $c_{t_np_n,n,i}^{tp,d}$, but their  
explicit forms
are unimportant except for $m_{\rm cr}$,
which is given by
\beqa
Z_F\ Z_m \ m_{\rm cr} a &=& c_{10,3}^{10,5} - c_{10,3}^{00,4}-\mu_0^2
a^2 \tilde c_{10,3}^{01,3}
\eeqa
where $\mu a\ \tilde c_{10,3}^{01,3} \equiv c_{10,3}^{01,3}$.
Using the selection rule (\ref{eq:rule2}) it is easy to show that
$c_{10,3}^{10,5}$,  $c_{10,3}^{00,4}$ and $\tilde c_{10,3}^{01,3}$
are even functions of $\mu_0 a$ .

Using renormalized fields (\ref{eq:redefine_field}) and parameters
(\ref{eq:redefine_para}), we finally obtain (\ref{eq:SymAction}) --
(\ref{eq:SymActionOdd}) in the main text.

%
\subsection{Symanzik expansion of operators}
%
Using the selection rules eqs.(\ref{eq:rule1}) and (\ref{eq:rule2}),
we here determine the structure of the Symanzik expansion for the  
lattice
operator, given by
\beqa
{\cal O}_{\rm lat}^{tp,d} &\Leftrightarrow & {\cal O}_{\rm eff}^{tp,d}
=\sum_{n=0}^\infty a^{n-d} \sum_{t_n,p_n}\sum_i c_{t_n p_n,n,i_n}^
{tp,d} {\cal O}
^{t_np_n,n}_{i_n}  .
\eeqa
In the case with $d=2 s$, the selection rule (\ref{eq:rule1}) gives
\beqa
{\cal O}_{\rm eff}^{tp,2s}
&=&\sum_{l=0}^\infty\sum_i \left[ a^{2(l-s)}  \left\{ c_{tp,2l,i}^{tp,
2s} {\cal O}^{tp,2l}_{i}
+ c_{\bar t\bar p,2l,i}^{tp,2s} {\cal O}^{\bar t\bar p,2l}_{i}\right\} 
\right.\nn\\
&&\hspace{1cm}+\left.
a^{2(l-s)+1}  \left\{ c_{\bar t p,2l+1,i}^{tp,2s} {\cal O}^{\bar t p,
2l+1}_{i}
+ c_{t\bar p,2l+1,i}^{tp,2s} {\cal O}^{t\bar p,2l+1}_{i}\right\}
\right]
\eeqa
where $\bar t = 1-t$ and $\bar p=1-p$.
Furthermore, using the second selection rule (\ref{eq:rule2}), we have
\beqa
{\cal O}_{\rm eff}^{tp,2s}
&=&\sum_{l=0}^\infty\sum_i \left[ a^{2(l-s)}  \left\{ c_{tp,2l,i}^{tp,
2s} {\cal O}^{tp,2l}_{i}
+ \mu_R\, \tilde c_{t\bar p,2l-1,i}^{tp,2s} {\cal O}^{t\bar p,2l-1}_{i}
\right\}
\right. \nn \\
&& \hspace{1cm}+ \left.
a^{2(l-s)+1}  \left\{ c_{\bar t p,2l+1,i}^{tp,2s} {\cal O}^{\bar t p,
2l+1}_{i}
+ \mu_R\,  c_{\bar t\bar p,2l,i}^{tp,2s} {\cal O}^{\bar t\bar p,2l}_{i}
\right\}
\right]  .
\eeqa
Similarly, for $d=2s+1$ we obtain
\beqa
{\cal O}_{\rm eff}^{tp,2s+1}
&=&\sum_{l=0}^\infty\sum_i \left[ a^{2(l-s)}  \left\{ c_{tp,2l+1,i}^
{tp,2s+1} {\cal O}^{tp,2l+1}_{i}
+ \mu_R\, \tilde c_{t\bar p,2l,i}^{tp,2s+1} {\cal O}^{t\bar p,2l}_{i}
\right\}
\right. \nn \\
&& \hspace{1cm}+  \left.
a^{2(l-s)+1}  \left\{ c_{\bar t p,2l,i}^{tp,2s+1} {\cal O}^{\bar t p,
2l}_{i}
+ \mu_R\,  c_{\bar t\bar p,2l-1,i}^{tp,2s+1} {\cal O}^{\bar t\bar p, 
2l-1}
_{i}\right\}
\right]  .
\eeqa
Combining the results
and rewriting the operators in terms of
renormalized fields we finally obtain eq.(\ref{eq:OP_expand2}).
%
\subsection{Expressions for $m_q={\rm O}(a)$ case}
%
In the case of $m_R ={\rm O}(a)$,
the expansion of  $e^{S_{\rm eff}}$ becomes
\beqa
e^{S_{\rm eff}}&=&e^{S_0} \exp\left\{m_R S_{m_R} +\sum_{n=1}^\infty
\left[
a^{2n} S_{2n}^0 + a^{2n-1} S_{2n-1}^1 \right] \right\} \nonumber \\
&\equiv & e^{S_0}\sum_{k,n=0}^\infty  \frac{m_R^k }{k!}S_{m_R}^{k}
a^n S^{(n)} .
\eeqa
Under the $T_1$ transformation, it is easy to see that
\beqa
T_1 :\ S_{m_R}^k &\rightarrow& (-1)^k S_{m_R}^k, \qquad
S^{(n)}\rightarrow  (-1)^n S^{(n)} .
\eeqa
Expanding both the action and the operator, and using the fact that
terms with $t+l+k+n =$ odd in the above expansion vanish
by the maximal twist condition (\ref{eq:maximal}),
we obtain
\beqa
\langle {\cal O}_{\rm eff, R,sub}^{pt,d} (\{x\} ) \rangle_{S_{\rm eff}}
&=& \delta_{t.0}\langle {\cal O}^{tp,d}_R\rangle_{S_0}\nn\\
&+&
\sum_{s=1}^\infty \sum_{k=0}^{2s-t} a^{2s-k-t} m_R^k  F_{d}^{2s-
t,k}(\{x\}, g_R^2,
\log(\Lambda a), \mu_R ; m_R a, \mu_R^2a^2),
\eeqa
where
\beqa
F_{d}^{2s-t,k}(\{x\}, g_R^2, \log(\Lambda a), \mu_R ; m_R a,
\mu_R^2a^2) &=& \sum_{l=0}^{2s-t-k}
\langle {\cal O}_{R;tpd}^{[l+t],d+l}(\{x\}) \frac{S_{m_R}^k}{k!} S^
{(2s-t-k-
l)} \rangle_{S_0} \nonumber \\
\eeqa
is an analytic function for small $m_R a$ and $\mu_R^2 a^2$.
This expression tells us that
\beqa
\langle {\cal O}_{\rm eff, R,sub}^{tp,d} (\{x\} ) \rangle_{S_{\rm eff}}
&=&\left\{
\begin{array}{ll}
\langle {\cal O}_R^{tp,d}\rangle_{S_0} + {\rm O}(a^2), & t=0 \\
{\rm O}(a), & t=1 \\
\end{array}\right.
\eeqa
for $m_R ={\rm O}(a)$.

If we take $m_R$ odd in $a$ such that $m_R = a f(a^2)$,
we have
\beqa
\langle {\cal O}^{tp,d}_{\rm eff, R,sub} (\{x\} ) \rangle_{S_{\rm eff}}
&=& \delta_{t.0}\langle {\cal O}^{tp,d}_R\rangle_{S_0}\nn\\
&+&
\sum_{s=1}^\infty a^{2s-t} F_d^{2s-t}(\{x\}, g_R^2, \log(\Lambda
a), \mu_R ;
a^2 , \mu_R^2a^2) \nonumber \\
&=&\left\{
\begin{array}{ll}
   \langle {\cal O}^{tp,d}_R\rangle_{S_0} + a^2 F_d^2 + a^4 F_d^4 +
\cdots, &  t=0 \\
aF_d^1 + a^3 F_d^3+\cdots, & t=1 \\
\end{array}\right.
\label{eq:main_res2}
\eeqa
where
\beqa
F_d^{2s-t}(\{x\}, g_R^2, \log(\Lambda a), \mu_R ; a^2 , \mu_R^2a^2) &=&
\sum_{k=0}^{2s-t}  (f(a^2))^k  F_{d}^{2s-t,k}(\{x\}, g_R^2, \log
(\Lambda a), \mu_R ; a^2 f^2, \mu_R^2a^2) .
\nonumber
\eeqa
%
\subsection{Maximal twist condition on the lattice}
%
The maximal twist condition on the lattice leads to
\beqa
0&=& \langle {\cal O}^{1p,d}_{\rm lat, R, sub}\rangle =
\langle {\cal O}^{1p,d}_{\rm eff, R, sub}\rangle =
\sum_{s=1}^\infty\sum_k^{2s-1} a^{2s-k-1}m_R^k F_d^{2s-1,k}\nn\\
&=& \sum_{k=0}^\infty \sum_{s=k+1}^\infty\left[
m_R^{2k} a^{2(s-k)-1} F_d^{2s-1,2k} + m_R^{2k+1} a^{2(s-k-1)} F_d^
{2s-1,2k+1}\right]\nn\\
&=& a H_d^0(\mu_R; a^2,m_R^2,m_R a, \mu_R^2a^2) + m_R
H_d^0(\mu_R; a^2,m_R^2,m_R a, \mu_R^2a^2),
\eeqa
where
\beqa
H_d^\delta(\mu_R; a^2,m_R^2,m_R a, \mu_R^2a^2) &\equiv&
\sum_{k=0}^\infty m_R^{2k} \sum_{s=k+1}^\infty a^{2(s-k-1)} F_d^
{2s-1,2k+\delta}
(g_R^2,\log(\Lambda a),\mu_R; m_R a, \mu_R^2a^2) \nonumber
\eeqa
for $\delta=0,1$,
and we keep the dependency on $g_R^2$ and $\log(\Lambda a)$ implicit in
$H_d^\delta$.

%
\section{Ward-Takahashi angle, pion mass and decay constant in WChPT}
\label{appendixC}
%
In this appendix we provide some details about the calculation of $
\cot\omega_{\rm WT}$ and the pseudo scalar masses and decay constant.
At leading order in our power counting scheme this has already be
done in Ref.\ \cite{Aoki:2004ta}, and we refer also to this reference.

Our first observable is the twist angle $\omega_{\rm WT}$ defined in
eq.\ \pref{Def:cotw}.
Instead of eq. \pref{Def:cotw} the twist angle can also be expressed as
\beqa\label{omegaWT}
\cot\omega_{\rm WT}&=& \frac{\langle \partial_\mu A_\mu^{2}\ P^{2}
\rangle}{\langle \partial_\mu V_\mu^1\ P^2\rangle}.
\eeqa
The extra derivative gives rise to an additional factor of the pion
mass in both the numerator and the denominator, which consequently
cancels.

The Noether's currents appearing in the correlators on the right hand side of
eq.\ \pref{omegaWT} are given by ($a=1,2$)
\bea
V_{\mu}^{a} & = & V_{0,\mu}^a \Big[1+\frac{c_{0}a}{4}\langle
\Sigma +\Sigma^{\dagger}\rangle \Big], \qquad
 V_{0,\mu}^a = \frac{if^{2}B}{4} \langle
\tau_{a}(\Sigma\partial_{\mu}\Sigma^{\dagger} + \Sigma^{\dagger}
\partial_{\mu}\Sigma )\rangle,\\
A_{\mu}^{a} & = & A_{0,\mu}^a \Big[1+\frac{c_{0}a}{4}\langle
\Sigma +\Sigma^{\dagger}\rangle \Big] ,\qquad
A_{0,\mu}^a=\frac{if^{2}B}{4}\langle
\tau_{a}(\Sigma\partial_{\mu}\Sigma^{\dagger} - \Sigma^{\dagger}
\partial_{\mu}\Sigma )\rangle.
\eea
The factor involving $c_{0}a$ stems from the wave function
renormalization due to the O($p^{2} a$) contribution in chiral
Lagrangian, cf.\ \pref{ChiralLag}.
In Ref.\ \cite{Aoki:2004ta} $\cowt$ was computed without the O($a
\mu,a^{3})$ contributions. Repeating the calculation including these
terms we find
\bea\label{Result:cotomega}
\cot\omega_{\rm WT} &=&  \frac{t}{\sqrt{1-t^{2}}}\,\,=\,\, \cot\phi\,.
\eea
This is the same result as in Ref.\ \cite{Aoki:2004ta}. The
functional dependence of $\cowt$ is unchanged, and the O($a\mu,a^{3})
$ corrections contribute only indirectly through the gap equation.

Result eq.\ \pref{Result:cotomega} assumes that the currents in the 
correlators are Noether's ones that stem from the vector and
axial vector Ward-Takahashi identities \cite{Frezzotti:2001ea}. This means that
the point split currents must be used in the lattice simulation.
However, local currents are often used instead of there point split
counter parts. This introduces additional contributions proportional
to $a$. Taking into account the leading corrections of O($a$) only we
obtain
\bea
Z_V V_{\mu}^{a, \rm local} & = & V_{0,\mu}^a\Big[ 1+ \frac{c_{\rm V}}{4}a
\langle \Sigma +\Sigma^{\dagger}\rangle\Big],
\\
Z_A A_{\mu}^{a, \rm local} & = & A_{0,\mu}^a\Big[ 1+ \frac{c_{\rm A}}{4}a
\langle \Sigma +\Sigma^{\dagger}\rangle\Big]+\frac{if^2B}{4}
\tilde c_A \partial_\mu\langle\Sigma-\Sigma^\dagger\rangle ,
\eea
where $c_{\rm V,A}$ and $\tilde c_A$
are additional coefficients parameterizing the
lattice artifacts stemming from the currents. Using these currents
the expression  \pref{Result:cotomega} changes to
\bea\label{Result:cotomegaLocal}
\cot\omega_{\rm WT} &=& \frac{Z_V}{Z_A}
\frac{1}{1+ a c_{\rm V}t}\frac
{(1+ a c_{\rm A}t)t-\tilde c_A a}{\sqrt{1-t^{2}}},
\eea
which is the result in Eq.\ \pref{cotw:local}.

In order to calculate the pion masses  we expand $\Sigma$ around the
vacuum configuration $\Sigma_{0}$ defined in \pref{AnsatzVEV}. We
parametrize the field $\Sigma$ in terms of the pion fields according
to\footnote{This  parameterization differs slightly from the one used
in Ref.\ \cite{Aoki:2004ta}. This difference does not affect any of
the results in this reference.}
\beqa
\Sigma(x) &=& \Sigma_{0}^{\frac{1}{2}} \exp \Big(\sum_{i=1}^{3}{i\pi_i
(x) \tau_i/f}\Big)\Sigma_{0}^{\frac{1}{2}}.
\eeqa
Using this form in expression \pref{PotentialEnergy2} for the
potential energy we expand in powers  of the field $\pi$. The
contribution quadratic in $\pi$ leads to the pion mass formulae
\beqa
m_{\pi_{\pm}}^2 &=& 2B m^{\prime}\cos(\phi-\omega_{L})+ 2W_{0}a \cos\phi
-2c_2a^{2}
\cos^2\phi\nn\\
& &+ 3c_{3}a^{3}\cos^{3}\phi
+2\tilde{c}_{2} a m^{\prime}\cos\phi \cos(\phi-\omega_{L}),  \\
m_{\pi_3}^2 &=& m_{\pi_a}^2+2c_2a^{2}\sin^2\phi- 6c_{3}a^{3} \sin^{2}
\phi\cos\phi- 2\tilde{c}_{2}a m^{\prime}\sin\phi\sin(\phi-\omega_{L})
.
\eeqa
Expressing this in terms of $m$ and $\mu$ and taking into account the
gap equation we can express the mass for the charged pion  
alternatively as
\beqa
m_{\pi_{ \pm}}^2 &=&\frac{2B\mu}{\sqrt{1-t^2}}\frac{1 + \beta_{\mu} t}
{ 1+c_0 a t}.\label{AlternativeMpi}
\eeqa 
Note that the expression \pref{AlternativeMpi} is not singular for
$t=1$. From the gap equation one can infer that $t=1$ can be a
solution only if $\mu=0$, and in that case the result  \pref
{AlternativeMpi} is not well defined. For $t\neq0$ one can, using the  
gap equation,
rewrite the pion mass formula as in \pref{Pionmass2},
which is well behaved for $t=1$.

The decay constant is conveniently computed with the so-called
indirect method \cite{Frezzotti:2001ea,Jansen:2003ir}, which is based
on an exact  PCVC relation  and does not require the computation of
any renormalization constants: 
\bea\label{Def:fpiInd}
f_{\pi} & = & \frac{2\mu}{\mpis} \langle 0| P^{\pm}|\pi_\mp\rangle \,.
\eea
In order to calculate the decay constant according to Eq.\ \pref
{Def:fpiInd} we need the matrix element of the pseudo scalar between
the vacuum and the one pion state, where the pseudo scalar in the
effective theory is defined by
($\tau_{\pm} = \frac{\tau_{1}\pm i\tau_{2}}{\sqrt{2}}$) 
\bea
P^{\pm}& =& \frac{f^{2}B}{4i}\Big[1+\frac{\betamu}{4}\langle\Sigma +
\Sigma^{\dagger}\rangle \Big] \langle
\tau_{\pm}(\Sigma -\Sigma^{\dagger})\rangle\,
\qquad \pi_\pm =\frac{\pi_1\pm i \pi_2}{\sqrt{2}}
\eea 
The matrix element in eq.\ \pref{Def:fpiInd} is readily calculated at
tree level with the result
\bea
\langle0| P^{\pm}|\pi_\mp\rangle & =& fB (1 + \betamu t).
\eea 
With the expression for the charged pion mass we obtain
\bea\label{Result:fpi}
f_{\pi}& =& f (1+c_{0}at ) \sqrt{1-t^{2}}
\eea
for the decay constant.
%

%
\end{appendix}
%

%

%

\begin{thebibliography}{10}
%
\bibitem{Frezzotti:2001ea}
R.~Frezzotti, S.~Sint and P.~Weisz,
\newblock JHEP {\bf 07} (2001) 048.

\bibitem{Frezzotti:2000nk}
R.~Frezzotti, P.~A. Grassi, S.~Sint and P.~Weisz,
\newblock JHEP {\bf 08} (2001) 058.

\bibitem{Bietenholz:2004wv}
W.~Bietenholz {\em et~al.},
\newblock JHEP {\bf 12} (2004) 044.

\bibitem{Abdel-Rehim:2005gz}
A.~M. Abdel-Rehim, R.~Lewis and R.~M. Woloshyn,
\newblock Phys. Rev. {\bf D71} (2005) 094505.

\bibitem{Jansen:2005gf}
K.~Jansen {\em et~al.},
\newblock Phys. Lett. {\bf B619} (2005) 184.

\bibitem{Jansen:2005kk}
K.~Jansen {\em et~al.},
\newblock JHEP {\bf 09} (2005) 071.

\bibitem{Urbach:2005ji}
C.~Urbach, K.~Jansen, A.~Shindler and U.~Wenger,
\newblock Comput. Phys. Commun. {\bf 174} (2006) 87.

\bibitem{Frezzotti:2003ni}
R.~Frezzotti and G.~C. Rossi,
\newblock JHEP {\bf 08} (2004) 007.

\bibitem{Frezzotti:2004wz}
R.~Frezzotti and G.~C. Rossi,
\newblock JHEP {\bf 10} (2004) 070.

\bibitem{Frezzotti:2005zm}
R.~Frezzotti and G.~Rossi,
\newblock hep-lat/0507030.

\bibitem{Shindler:2005vj}
A.~Shindler,
\newblock PoS {\bf LAT2005} (2005) 014.

\bibitem{Aoki:2004ta}
S.~Aoki and O.~B{\"a}r,
\newblock Phys. Rev. {\bf D70} (2004) 116011.

\bibitem{Frezzotti:2005gi}
R.~Frezzotti, G.~Martinelli, M.~Papinutto and G.~C. Rossi,
\newblock hep-lat/0503034.

\bibitem{Jansen:2005mf}
K.~Jansen {\em et~al.},
\newblock PoS {\bf LAT2005} (2005) 231.

\bibitem{Sharpe:1998xm}
S.~R. Sharpe and J.~Singleton, Robert,
\newblock Phys. Rev. {\bf D58} (1998) 074501.

\bibitem{Rupak:2002sm}
G.~Rupak and N.~Shoresh,
\newblock Phys. Rev. {\bf D66} (2002) 054503.

\bibitem{Bar:2004xp}
O.~B{\"a}r,
\newblock Nucl. Phys. Proc. Suppl. {\bf 140} (2005) 106.

\bibitem{Sharpe:2004ps}
S.~R. Sharpe and J.~M.~S. Wu,
\newblock Phys. Rev. {\bf D70} (2004) 094029.

\bibitem{Symanzik:1983dc}
K.~Symanzik,
\newblock Nucl. Phys. {\bf B226} (1983) 187.

\bibitem{Symanzik:1983gh}
K.~Symanzik,
\newblock Nucl. Phys. {\bf B226} (1983) 205.

\bibitem{Aoki:1983qi}
S.~Aoki,
\newblock Phys. Rev. {\bf D30} (1984) 2653.

\bibitem{Aoki:1985mk}
S.~Aoki,
\newblock Phys. Rev. {\bf D33} (1986) 2399.

\bibitem{Aoki:1986kt}
S.~Aoki,
\newblock Phys. Rev. {\bf D34} (1986) 3170.

\bibitem{Aoki:1986xr}
S.~Aoki,
\newblock Phys. Rev. Lett. {\bf 57} (1986) 3136.

\bibitem{Aoki:1987us}
S.~Aoki,
\newblock Nucl. Phys. {\bf B314} (1989) 79.

\bibitem{SintNara}
S.~Sint,
\newblock Lectures given at the Nara workshop, Oct. 31 -- Dec. 11, 2005, Nara,
  Japan. To be published.

\bibitem{Bar:2003xq}
O.~B{\"a}r, G.~Rupak and N.~Shoresh,
\newblock Nucl. Phys. Proc. Suppl. {\bf 129} (2004) 185.

\bibitem{Aoki:2003yv}
S.~Aoki,
\newblock Phys. Rev. {\bf D68} (2003) 054508.

\bibitem{Munster:2003ba}
G.~M{\"u}nster and C.~Schmidt,
\newblock Europhys. Lett. {\bf 66} (2004) 652.

\bibitem{Sharpe:2004ny}
S.~R. Sharpe and J.~M.~S. Wu,
\newblock Phys. Rev. {\bf D71} (2005) 074501.

\bibitem{Sharpe:2005rq}
S.~R. Sharpe,
\newblock Phys. Rev. {\bf D72} (2005) 074510.

\bibitem{Munster:2004am}
G.~M{\"u}nster,
\newblock JHEP {\bf 09} (2004) 035.

\bibitem{Aoki:2005ii}
S.~Aoki and O.~B{\"a}r,
\newblock PoS {\bf LAT2005} (2005) 046.

\bibitem{Sharpe:2004bv}
S.~R. Sharpe and J.~M.~S. Wu,
\newblock Nucl. Phys. Proc. Suppl. {\bf 140} (2005) 323.

\bibitem{Bar:2003mh}
O.~B{\"a}r, G.~Rupak and N.~Shoresh,
\newblock Phys. Rev. {\bf D70} (2004) 034508.

\bibitem{Gasser:1983yg}
J.~Gasser and H.~Leutwyler,
\newblock Ann. Phys. {\bf 158} (1984) 142.

\bibitem{Gasser:1984gg}
J.~Gasser and H.~Leutwyler,
\newblock Nucl. Phys. {\bf B250} (1985) 465.

\bibitem{Bar:2002nr}
O.~B{\"a}r, G.~Rupak and N.~Shoresh,
\newblock Phys. Rev. {\bf D67} (2003) 114505.

\bibitem{Aoki:2005mb}
S.~Aoki, O.~B{\"a}r, S.~Takeda and T.~Ishikawa,
\newblock Phys. Rev. {\bf D73} (2006) 014511.

\bibitem{ABT}
S.~Aoki, O.~B{\"a}r and S.~Takeda,
\newblock Currents and densities in Wilson Chiral Perturbation theory, in
  preparation.

\bibitem{SharpeNara}
S.~R. Sharpe,
\newblock Lectures given at the Nara workshop, Oct. 31 -- Dec. 11, 2005, Nara,
  Japan. To be published.

\bibitem{Farchioni:2004us}
F.~Farchioni {\em et~al.},
\newblock Eur. Phys. J. {\bf C39} (2005) 421.

\bibitem{Jansen:2003ir}
K.~Jansen, A.~Shindler, C.~Urbach and I.~Wetzorke,
\newblock Phys. Lett. {\bf B586} (2004) 432.

\end{thebibliography}
\end{document}